\documentclass[a4paper,11pt]{article}
\pdfoutput=1
 \usepackage{jheppub} 
 \usepackage{graphicx}
\usepackage{amsmath}
 \usepackage{amssymb}
 \usepackage{slashed}
\usepackage{dcolumn}% Align table columns on decimal point
\usepackage{bm}% bold math 
\usepackage{color}
\usepackage{multirow}

\allowdisplaybreaks

 \newcommand{\lsim}{{\;\raise0.3ex\hbox{$<$\kern-0.75em\raise-1.1ex\hbox{$\sim$}}\;}}
\newcommand{\gsim}{{\;\raise0.3ex\hbox{$>$\kern-0.75em\raise-1.1ex\hbox{$\sim$}}\;}}
\def\bea{\begin{eqnarray}}
\def\eea{\end{eqnarray}}
\def\bec{\begin{center}}
\def\ec{\end{center}}

\def\beq{\begin{equation}}
\def\eeq{\end{equation}}

\def\bea{\begin{eqnarray}}
\def\eea{\end{eqnarray}}
\def\beq#1\eeq{\begin{align}#1\end{align}}
\def\beqnn#1\eeq{\begin{align*}#1\end{align*}}
\def\ba{\begin{array}}
\def\ea{\end{array}}
\def\bc{\begin{center}}
\def\ec{\end{center}}

\preprint{CTPU-PTC-19-27,\,KIAS-P19057}

\title{Axion scales and couplings with St\"uckelberg mixing}
 
 \author{
    Kiwoon Choi$^{a}$\footnote{Electronic address: kchoi@ibs.re.kr},
    Chang Sub Shin$^{a}$\footnote{Electronic address: csshin@ibs.re.kr},
    Seokhoon Yun$^{b}$\footnote{Electronic address: SeokhoonYun@kias.re.kr}
    }
     
\affiliation{
 $^a$Center for Theoretical Physics of the Universe,  Institute for Basic Science, Daejeon 34126,\\
  South Korea \\
 $^b$Korea Institute for Advanced Study, Seoul 02455,  South Korea 
    }

\abstract{
We study the axion field range  and low energy couplings in models with  St\"uckelberg mixing 
between axions and $U(1)$ gauge bosons.
It is noted that the gauge-invariant axion combination $\xi$ in the model is periodic {\it modulo} an appropriate shift of gauge-variant axions eaten by the massive $U(1)$ gauge bosons, which in some cases makes the connection between the field range and the low energy couplings less transparent. 
   We  derive the field range of $\xi$  for generic forms of the axion kinetic metric  and $U(1)$ charges, and identify the field basis for which all non-derivative couplings of $\xi$ are quantized in a manner manifestly consistent with the periodicity  of $\xi$.
Generically St\"uckelberg mixing reduces the axion field range.  In particular, the mixings between $N$ axions and $(N-1)$ $U(1)$ gauge bosons typically result in an exponentially reduced field range $M_\xi ={\cal O}\left({k^{-(N-1)} f}/{\sqrt{N!}}\right)$  for the residual gauge-invariant axion $\xi$ in the limit $N\gg 1$
, where
$f$ and  $k$ denote the typical decay constant and the root mean square  of the $U(1)$ gauge charges of the original $N$ axions. 
Using simple examples, we study also the reparameterization-invariant physical quantities such as the axion effective potential and 1PI couplings to gauge bosons, which are determined by
the reparameterization-dependent axion couplings in the model.
}

\vspace{4cm}

\begin{document} 
\maketitle
\flushbottom

\section{Introduction}

Axions (or axion-like particles) are considered to be one of the most compelling candidates for physics beyond the Standard Model of particle physics~\cite{Kim:2008hd}. Axions are periodic scalar fields and much of their low energy physics are determined by the mass scale $M_a$ called the axion decay constant which defines the field range of the canonically normalized axion as 
\bea
a\equiv a+2\pi M_a.\eea
There have been a variety of different axions introduced so far in particle physics and cosmology, and the favoured range of the decay constant of those axions differ by many orders of magnitudes. In most cases, $M_a$ is considered to be well below the Planck scale $M_{\rm Pl}$~\cite{Kim:2008hd}, however in some cases it needs to be comparable to or even bigger than  $M_{\rm Pl}$~\cite{Freese:1990rb,Choi:1999xn,Graham:2015cka}. 
In theoretical side, it has been known for many years that potentially realistic string compactifications provide multiple axions
whose decay constants are often of the order of ${g^2M_{\rm Pl}}/{8\pi^2}$, where $g$ is the gauge coupling in the model~\cite{Witten:1984dg,Choi:1985je,Svrcek:2006yi,Arvanitaki:2009fg}. There  has been also an argument called the weak gravity conjecture on axions~\cite{ArkaniHamed:2006dz}, implying that $M_a\lesssim {\cal O}({g^2M_{\rm Pl}}/{8\pi^2})$ within a theory defined at the scale of quantum gravity. Motivated by these, various mechanisms
have been proposed to widen the possible range of $M_a$ in the low energy effective theory
%in particular to enhance $f_a$ to a possibly trans-Planckian value 
 starting from a UV theory  whose axion scales are limited to be within certain range~\cite{ArkaniHamed:2003wu,Kim:2004rp,Bachlechner:2014gfa,Dimopoulos:2005ac,Junghans:2015hba,Silverstein:2008sg,Kaloper:2008fb,
 Marchesano:2014mla,Choi:2014rja,Higaki:2014pja,Bachlechner:2014hsa,Ben-Dayan:2014zsa,delaFuente:2014aca,Shiu:2015uva,Shiu:2015xda,Brown:2015iha,Hebecker:2015rya,Choi:2015fiu,Kaplan:2015fuy,Giudice:2016yja,Heidenreich:2015wga,Shiu:2018unx,Fonseca:2019aux,Montero:2015ofa,Bachlechner:2017zpb,Bachlechner:2017hsj,Rudelius:2014wla}

In this paper, we wish to revisit one of such mechanisms, utilizing the St\"uckelberg mixing between axions and $U(1)$ gauge bosons\footnote{In fact, all of our discussions are applicable also to the case that $U(1)$ gauge symmetries are broken by the conventional Higgs mechanism. The only difference between the Higgs mechanism and the St\"uckelberg mechanism is the existence of the radial partner of the  Goldstone bosons eaten by the $U(1)$ gauge bosons, which is not relevant for low energy physics that we are concerned with.} \cite{Shiu:2015xda,Shiu:2015uva,Fonseca:2019aux}.
%Here we use the St\"uckelberg mechanism without such radial partner for simplicity.}. 
%theories with $N$ axions in which $M<N$ combinations of axions are appropriately $U(1)$ gauge-charged to be absorbed in the %longitudinal components of the $M$ massive $U(1)$ gauge bosons through the  the St\"uckelberg mixingIt has been argued that %the field range and low energy couplings of the remained $(N-M)$ gauge-invariant axions 
It has been noticed  in \cite{Shiu:2015xda,Shiu:2015uva} that in some parameter limit of the St\"uckelberg mixing,
the coupling $1/f_\xi$ of the canonically normalized gauge-invariant axion combination $\xi$ 
%which is free from the St\"uckelberg mixing 
to non-Abelian gauge fields, i.e. $\frac{1}{32\pi^2}\frac{\xi}{f_\xi} G^{a\mu\nu}\tilde G^a_{\mu\nu}$, can be significantly smaller  than the inverse of the mass scales introduced in the UV theory.
Then, based on the expectation that $f_\xi$
is comparable to the axion decay constant $M_\xi$ which is defined by the axion periodicity $\xi\equiv \xi +2\pi M_\xi$, such suppression of $1/f_\xi$ was interpreted as an indication of the enhanced axion field range in the corresponding parameter limit.
Recently the possibility of enhanced axion field range   has been explored again with a simple model yielding  an exponentially suppressed $1/f_\xi$ in a natural manner~\cite{Fonseca:2019aux}.
% which isdue to a scale hierarchy generated by either warped extra dimension or nearly conformal 4D dynamics.  
On the other hand, in the presence of gauge-charged fermions, the axion coupling $1/f_\xi$ varies under the $\xi$-dependent phase rotation of fermion fields, while
the axion field range $M_\xi$ is invariant under such field redefinition. In such case,
the connection between the coupling $1/f_\xi$ and the axion decay constant $M_\xi$ depends on the choice of field basis, so deserves more careful analysis.
%Indeed, in axion models with the St\"uckelberg mixing,
%quite often  $f_\xi$ is
%hierarchically  bigger than $M_\xi$ in the field basis chosen in the UV theory.

Recently the authors of \cite{Bonnefoy:2018ibr}
% on the St\"uckelberg mixing between axions and $U(1)$ gauge bosons,
examined a model with  clockwork-type St\"uckelberg mixings between $N$ axions and $(N-1)$ $U(1)$ gauge bosons,
%$U(1)$ charges of the axions for the St\"uckelberg mixings with multiple $U(1)$ gauge bosons,
and noticed that  $\xi$ is so well protected by 
% $\frac{1}{32\pi^2}\frac{\xi}{f_\xi} G^{a\mu\nu}\tilde G^a_{\mu\nu}$ should involve a large coefficient to be 
%compatible with 
the $(N-1)$ $U(1)$ gauge symmetries from getting a mass in the limit $N\gg 1$.
% so can be ultra-light in a natural manner. 
As we will see, such protection of $\xi$ from being massive is deeply connected with the exponential reduction of the axion field range $M_\xi$ by the St\"uckelberg mixing in the limit $N\gg 1$.
%reduce the axion field range to a value much smaller 
%the mass scales in the underlying UV theory.  
%This motivates us to study how the  St\"uckelberg mixing affects the axion scales and low energy couplings in a general %framework.
Therefore  the previous studies suggest that  St\"uckelberg mixing between axions and $U(1)$ gauge bosons can result in rich consequences in low energy axion physics. 
We wish to examine those consequences 
%between axions and $U(1)$ gauge bosons
in a general framework which can cover all of the previous studies, while clarifying some confusions
made in the previous works. 

The organization of this paper is as follows. In the next section, we discuss the axion field range and low energy couplings
in generic axion models with  St\"uckelberg mixing. 
 We first note that in such models the gauge-invariant axion combination $\xi$ is periodic {\it modulo} a  shift of the gauge-variant axion combinations eaten by the massive $U(1)$ gauge bosons, which is determined by the
kinetic metric and $U(1)$ gauge charges of the original axions. This often
makes the connection between the field range and low energy couplings of $\xi$ less transparent. 
  % making the connection between its field range and low energy couplings less transparent. 
 % or modulo an equivalent $U(1)$ transformation of gauge-charged matter fields.
%This makes the connection between the axion field range $\Delta\xi$ and the axion non-derivative couplings such as   $\frac{1}{32\pi^2}\frac{\xi}{f_\xi} G^{a\mu\nu}\tilde G^a_{\mu\nu}$ is more involved.
% This is in fact not unexpected as 
% of $\xi$ to matter and gauge fields are varying under the $\xi$-dependent redefinition of the gauge-charged matter fields.
 We then derive the field range of $\xi$ for generic forms of the axion kinetic metric  and $U(1)$ charges, 
   %We note that in some cases the axion coupling to non-Abelian gauge fields,  $\xi G^{a\mu\nu}\tilde G^a_{\mu\nu}$, can have an %unambiguous connection  to the axion field range $\Delta\xi$ {\it only} when the coupling is shifted by an appropriate field %redefinition involving $\xi$-dependent $U(1)_A$ transformation of the gauge-charged fermion fields. 
and discuss the axion couplings to matter and gauge fields, which depend on the choice of the matter field basis.
We also identify the field basis for which all non-derivative couplings of $\xi$ are quantized  in a manner manifestly consistent with the axion  periodicity  $\xi\equiv \xi +2\pi M_\xi$. 
%the corresponding coupling of $\xi$ to  $G^{a\mu\nu}\tilde G^a_{\mu\nu}$ is not %manifestly consistent with the axion periodicity $\xi\equiv \xi+\Delta\xi$,
% but can be made to be consistent when the coupling is shifted  by 
%generically has a continuous value in the unit of $2\pi/\Delta\xi$, so by itself does not provide any information on the %field range $\Delta\xi$. Yet one can make an appropriate  field redefinition, which corresponds to a $\xi$-dependent $U(1)_A$ %transformation of the matter fields in the model, after which all non-derivative couplings of $\xi$, including the coupling %to the YM instantons, are integer-multiples of $2\pi/\Delta\xi$, and therefore manifestly consistent with the axion %periodicity $\xi\equiv \xi+\Delta\xi$. 
It is noted also that St\"uckelberg mixing typically reduces the axion field range to a value smaller than the mass scales in the UV theory. In particular,
for the case of St\"uckelberg mixing between   $N$ axions and $(N-1)$ $U(1)$ gauge bosons,
the axion field range is reduced as $M_\xi ={\cal O}(k^{-(N-1)}f/\sqrt{N!})$ in the limit $N\gg 1$, where $f$ and $k$ denote the typical decay constant and the root mean square  of the $U(1)$ gauge charges of the original $N$ axions.

In Sec.~\ref{sec:impl}, we apply the results of Sec.~\ref{sec:gen} to  specific examples to see the implications of our results. We first consider an illustrative simple  model of St\"uckelberg mixing between two axions and a single $U(1)$ gauge boson.
For this model, we study
 the reparameterization-invariant physical quantities such as the axion field range, axion 1PI amplitude to gauge bosons, and the axion effective potential induced by non-perturbative gauge dynamics, which are determined by the reparameterization-dependent  axion couplings in the model. Another example is the model studied in 
\cite{Bonnefoy:2018ibr}, involving  $N$ axions which have a clockwork-type St\"uckelberg mixing with $(N-1)$ $U(1)$ gauge bosons. For this model, we examine the field range and low energy couplings of
the gauge-invariant axion combination $\xi$, and discuss how much non-trivial it is to
generate an effective potential of $\xi$ in the limit $N\gg 1$.
  Sec.~\ref{sec:con} is the conclusion.

\section{Axion field range and couplings with  St\"uckelberg mixing}\label{sec:gen}

\subsection{St\"uckelberg mixing between two axions and single $U(1)$ gauge boson}
In this section, we examine the axion field range and low energy couplings
in generic axion models with the St\"uckelberg mixing. For simplicity,
we start with the case of two axions which have a St\"uckelberg mixing with single $U(1)_A$ gauge boson.
% which absorbs one combination of axions as its longitudinal component. 
%Such model has been discussed in \cite{} as an illustrative example to examine the effects of the St\"uckelberg mixings on the %axion field range and couplings  in the low energy effective theory.  
In addition to $U(1)_A$, the model involves also a non-Abelian gauge symmetry which will be chosen to be $SU(N_c)$ in the following discussion.   At high scales above the St\"uckelberg mass, the lagrangian density is given by
\bea
\label{model1}
{\cal L}&=&\frac{1}{2} \sum_{ij}G_{ij}\left(\partial_\mu\theta^i -k^iA_\mu\right)\left(\partial^\mu \theta^j -k^jA^\mu\right)-\frac{1}{4 g^2}F_{\mu\nu}F^{\mu\nu} 
-\frac{1}{4g_a^2} G^a_{\mu\nu} G^{a\mu\nu}\nonumber \\
&&+\,  \frac{1}{32\pi^2}\Big(\sum_i r_i\theta^i\Big) G^a_{\mu\nu}\tilde G^{a\mu\nu}
+ \frac{1}{32\pi^2} \Big(\sum_i s_i\theta^i\Big) F_{\mu\nu}\tilde F^{\mu\nu} \nonumber \\
&&+\, 
\sum_P |D_\mu\phi_P|^2 + 
\sum_I\bar\psi_I  i  \bar\sigma^\mu D_\mu  \psi_I 
 - \left(\mu_{IJ}e^{i\sum_i n^{IJ}_ i\theta^i}\psi_I\psi_J+{\rm h.c.}\right)\nonumber \\
 &&
 -\, \left(\lambda_{IJP} e^{i\sum_i n^{IJP}_i \theta^i}\phi_P\psi_I\psi_J+{\rm h.c.}\right)+\sum_i (\partial_\mu\theta^i-k^iA_\mu)J^\mu_i+ \cdots,
\eea 
where $\theta^i$ ($i=1,2$) are dimensionless axion fields normalized to have the $2\pi$ periodicity: 
\bea
\theta^i\equiv \theta^i+2\pi,\eea
 $A_\mu$ and $G_\mu^a$ denote the $U(1)_A\times SU(N_c)$ gauge fields, $\psi_I$ and $\phi_P$ are chiral fermions and complex scalar fields in the model, $J^\mu_i$ are gauge-invariant currents made of matter fields $\Phi=(\phi_P,\psi_I)$,
and the ellipsis stands for possible additional terms including the gauge-invariant potential of $\theta^i$ and $\phi_P$. We assume that 
the (approximate) continuous shift symmetries  $\theta^i\rightarrow \theta^i+{\rm constant}$ are good enough, so that the axion kinetic metric $G_{ij}$ is independent of $\theta^i$. 
Under  $U(1)_A$, the fields transform as
 \bea 
 \label{u1a}
U(1)_A:\ &&
A_\mu \to A_\mu  +\partial_\mu\Lambda, \quad 
\theta^i\to\theta^i + k^i\Lambda,\quad
\Phi\to e^{-iq_\Phi \Lambda}\,\Phi \,\,\, \left(\Phi=\psi_I, \phi_P\right), 
%&& {\cal O}_n(\psi_I,...)\to e^{-\sum_i n_ik_i\Lambda}{\cal O}_n(\psi_I,...),
\eea
where the $U(1)_A$ gauge transformation function $\Lambda(x)$ obeys  the periodicity condition 
\bea\Lambda(x)\equiv \Lambda(x)+2\pi,\eea
which ensures that the $U(1)_A$ charges $k^i$ and $q_\Phi$ have integer values.
%Note that the $U(1)_A$ transformation of the composite operator ${\cal O}_n(\psi_I,...)$ is 
The axion $\theta^i$ can have a variety of non-derivative couplings to the gauge and matter fields, some of which are
explicitly given and parametrized  by  $r_i, s_i, n^{IJ}_i$ and $n^{IJP}_i$ in (\ref{model1}), as well as the derivative couplings to the currents $J_i^\mu$. 
Here we choose the field basis for which the $2\pi$ periodicity of $\theta^i$ is {\it manifest}, i.e. the model is invariant under the discrete gauge symmetries
\bea
\label{shift_gauge}
\mathbb{Z}_i:  \,\,\, \theta^i\, \to\,\, \theta^i +2\pi  \quad (i =1,2),\eea
under which {\it only} $\theta^i$ transforms, while all other fields are invariant\footnote{Generically the discrete symmetry $\mathbb{Z}_i$ may include additional  transformations of light fields in the model, e.g. $\Phi\rightarrow e^{i\Delta_\Phi}\Phi$ for matter fields $\Phi=\{\phi_P,\psi_I\}$, as well as a change of discrete quantum numbers to define the effective theory (\ref{model1}), e.g. a shift of background flux which originates from the underlying UV theory.  The transformation $\Phi\rightarrow e^{i\Delta_\Phi}\Phi$ can be eliminated by making the $\theta^i$-dependent field redefinition: $\Phi\rightarrow e^{-i\Delta_\Phi\theta^i/2\pi}\Phi$, after which  $\Phi$ becomes invariant under $\mathbb{Z}_i$, while the lagrangian density is accordingly modified in the new field basis. As for the possibility of background flux which has a non-trivial transformation under $\mathbb{Z}_i$, if such flux exists, the $U(1)_A$-invariant axion combination can get a heavy mass from the flux together with the monodromy feature associated with the shift of flux \cite{Silverstein:2008sg,Kaloper:2008fb}. Here we are  interested in the effects of the St\"uckelberg mixing on low energy axion physics, and therefore consider the case without such background flux.}.   
In such field basis, the non-derivative coupling parameters  $r_i,s_i,n^{IJ}_i$ and $n^{IJP}_i$ have integer values\footnote{
%As there is no instanton-like configuration for $U(1)$ gauge fields, the parameters $s_i$ might be allowed to have a %continuous value. This point is not relevant for our discussion, so we simply assume that $s_i$ also have integer values 
%like $r_i, n^{IJ}_i$ and $n^{IJP}_i$.
Note that such quantization of non-derivative couplings of $\theta^i$ is based on the assumption
 that the lagrangian (\ref{model1}) is valid over the entire range of the axion fields $\theta^i$, which we take in this paper.  In some case, for instance the QCD axion $a_{\rm QCD}$ at scales below the QCD scale, the QCD mesons have non-trivial axion-dependent tadpoles, rendering 
the axion effective lagrangian valid over the full range of $a_{\rm QCD}$ to have a  complicate form. In such case,  one usually considers an effective lagrangian of small axion fluctuation $\delta a_{\rm QCD}$ around the vacuum, whose non-derivative couplings are not constrained to be quantized  in the unit of $1/\Delta a_{\rm QCD}$. }, which ensures that the model (\ref{model1}) is manifestly invariant under $\prod_i \mathbb{Z}_i$.
%We assume also that the lagrangian (\ref{model1}) is valid over the entire axion moduli space and the $U(1)_A$ gauge %transformation (\ref{u1a}) is well defined over the full range of $\Lambda(x)$. 
Obviously the fermion mass parameter  $\mu_{IJ}$ and the Yukawa coupling $\lambda_{IJP}$ can be nonzero  only when the corresponding operators are
invariant under  $SU(N_c)$, and  also satisfy the following $U(1)_A$ invariance conditions: 
\bea
\label{u1_condition}
 q_I+q_J=\sum_i n^{IJ}_i k^i , \quad q_I+q_J+q_P=\sum_i n^{IJP}_ik^i.\eea
We consider the case that the fermions $\{\psi_I\}$  form a vector-like representation of $SU(N_c)$, but can be  chiral under $U(1)_A$. Then there can be nonzero $[U(1)_A]^3$ and $U(1)_A\times [SU(N_c)]^2$ gauge anomalies, which should be  cancelled by the $U(1)_A$ variation of the axion couplings $\theta^i F\tilde F$ and $\theta^iG\tilde G$. This  requires  \bea
\label{gs_condition}
\sum_I q_I^3 +\sum_i s_ik^i= 2\sum_I q_I{\rm Tr}(T_a^2(\psi_I)) +\sum_i r_ik^i=0,\eea
where $T_a(\psi_I)$ denotes the $SU(N_c)$ generator for the fermion field $\psi_I$. 
 
For our subsequent discussion,  it is useful to define a complete set of integer-valued vectors and dual vectors in $\theta$-space, for which the integer-valued components of each vector are relatively prime.
%  i.e.  the greatest common divisor of $\chi_i$s (gcd($\vec\chi$))   equals $1$. 
One such vector is provided by the $U(1)_A$ charges of $\theta^i$ as
%the vector $\vec k_r$ can be defined as
\bea 
\vec k_{r} = \left(  k_r^1,  k_r^2 \right) \equiv \frac{\left(  k^1,  k^2 \right)}{{\rm gcd}(\vec k)} ,\eea
where ${\rm gcd}(\vec k)$ is the greatest common divisor of $k^1$ and $k^2$. 
We can construct the other linearly independent vector $\vec\ell = (\ell^1, \ell^2)$ 
and also the dual vectors $\vec{\tilde k} =(\tilde k_1, \tilde k_2)$ and $
\vec{\tilde \ell}=(\tilde \ell_1, \tilde \ell_2)$ from the conditions:
\bea \label{vec_dual}
\vec {\tilde k}\cdot \vec k_r =0,\quad 
\vec{\tilde k}\cdot \vec \ell = 1,\quad 
\vec{\tilde \ell}\cdot \vec \ell = 0,\quad
\vec{\tilde \ell}\cdot \vec k_r = 1. 
\eea 
For a given $\vec k$, the
 above conditions uniquely (up to sign) fix $\vec{\tilde k}$ as  
\bea 
\vec {\tilde k} =\pm (k_r^2, - k_r^1)= \pm\frac{(k^2, - k^1)}{{\rm gcd}(\vec k)}, 
\eea 
while $\vec\ell$ and $\vec{\tilde\ell}$ have additional degeneracy.  For $\vec\ell$ and $\vec{\tilde\ell}$ satisfying (\ref{vec_dual}), one easily finds    
\bea \label{degenerate}
\vec\ell^\prime =\vec \ell + q \vec k_r,\quad 
\vec{\tilde \ell}^{\prime} = \vec{\tilde \ell} - q \vec{\tilde k}
\eea 
are also a solution, where  $q$ is an arbitrary integer. At any rate,
 all solutions of (\ref{vec_dual}) satisfy the identity 
\bea \label{complete}
k_r^i \tilde \ell_j + \ell^i \tilde k_j = \delta^i_j 
\eea 
which turns out to be quite useful for our subsequent discussions.
As we will see in the later part of this section, the above construction of the integer-valued vectors ($\vec k_r, \vec \ell$) and dual vectors ($\vec{\tilde k}, \vec{\tilde \ell})$ 
can be easily 
generalized to the more general case of  $N (>2)$ axions which have the St\"uckelberg mixings with $(N-1)$  $U(1)$ gauge bosons.

In the above model, the $U(1)_A$ gauge boson gets a nonzero mass $gM_A$ through the St\"uckelberg mechanism, where
\bea
M_A^2(G,\vec k) = \sum_{ij}G_{ij}k^ik^j.\eea
%which will be assumed to be heavy enough in the subsequent discussions.
It is then straightforward to rewrite the axion kinetic terms in terms of the gauge-invariant physical axion $\xi$ and the gauge-variant $\zeta$ eaten by the massive $U(1)_A$ gauge boson:
\bea
\frac{1}{2} \sum_{ij}G_{ij}(\partial_\mu\theta^i -k^iA_\mu)(\partial^\mu \theta^j -k^jA^\mu)
=\frac{1}{2}(\partial_\mu\xi)^2+
\frac{1}{2}M_{A}^2\left(A_\mu-{\partial_\mu\zeta}\right)^2, \eea
where
\bea
\label{zeta_xi}
\xi  = M_\xi 
%\frac{1}{\sqrt{\sum_{ij}(G^{-1})_{ij}\tilde k_i\tilde k_j}}
\sum_i
\tilde k_i \theta^i  ,\quad  
\zeta=
%\frac{1}{\sqrt{\sum_{ij}G_{ij}k_ik_j}}
M_A^{-2}  \sum_{ij} G_{ij}k^i \theta^j 
 \eea
for
\bea
\label{M_xi}
 M_\xi(G,\vec k) =\frac{1}{\sqrt{\sum_{ij}(G^{-1})^{ij}\tilde k_i\tilde k_j}}.
\eea
Note that 
%\bea
%U(1)_A: \quad \zeta \to \, \zeta _\Lambda(x), \quad \xi \to \, \xi.\eea
we use the convention that the gauge-variant $\zeta$ is dimensionless, while the gauge-invariant $\xi$  is a canonically normalized field with mass dimension one.

\begin{figure}[t] %htbp]
\begin{center}
\includegraphics[width=7cm]{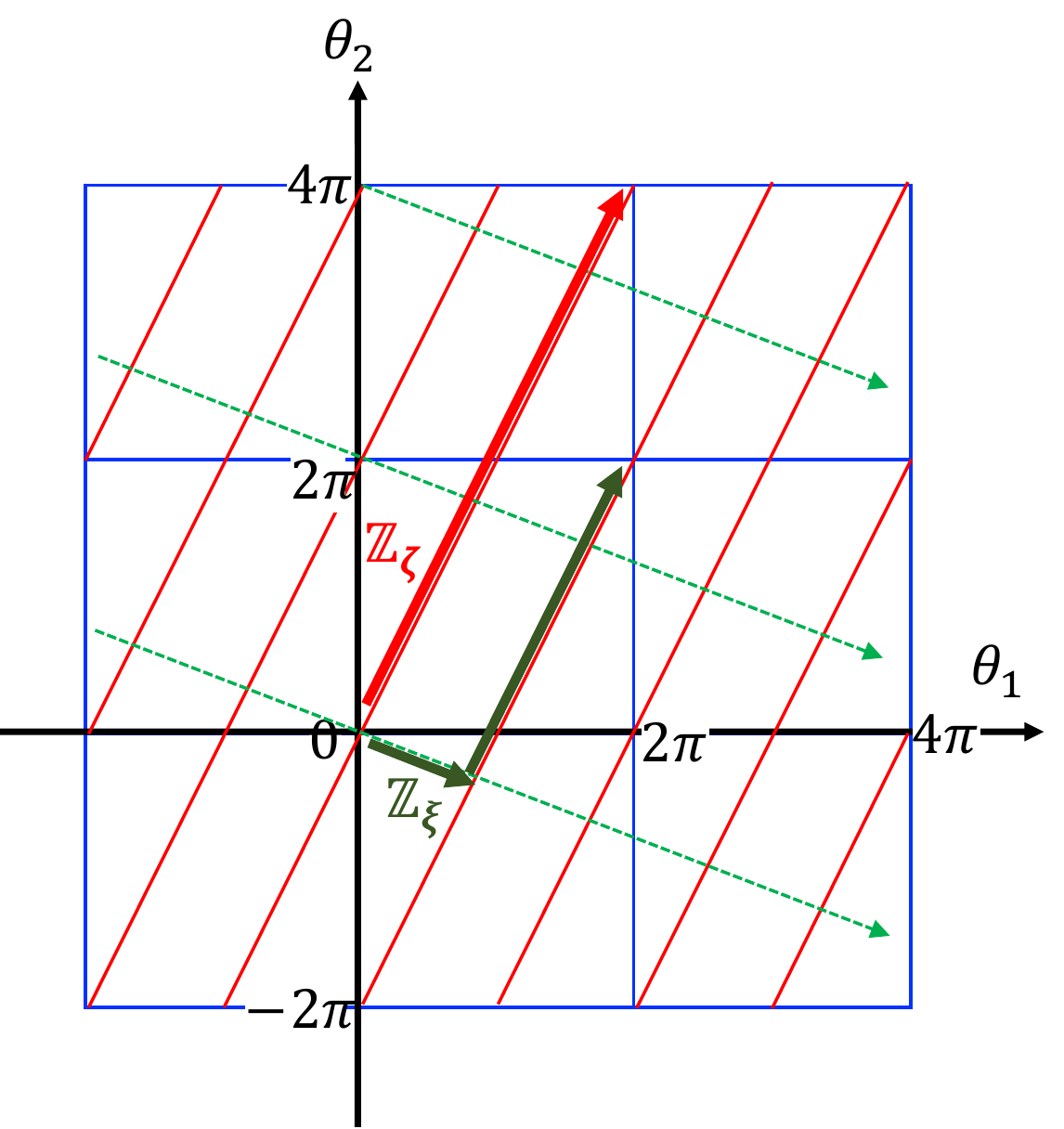}
\caption{Illustration of the  discrete gauge transformations $\mathbb{Z}_\zeta$ (thick red arrow) and $\mathbb{Z}_\xi$ (thick black arrow) in the moduli space of $\theta^i$. 
As an example, here we take $k_1= 1$ and $k_2=2$. The red lines are the coordinate axis of
 $\zeta$, while the green lines are the coordinate axis of $\xi$.  Note that $\partial\theta^i/\partial\zeta=k^i$ are integers, while $\partial\theta^i/\partial\xi$ are kinetic-metric-dependent continuous numbers, so
the two coordinate axes are not orthogonal to each other.
%Under $\mathbb{Z}_\zeta$ (the thick red line), $\zeta \to \zeta + 2\pi$. 
%For $\mathbb{Z}_\xi$ (the thick green line), $\xi/M_\xi \to 
%\xi/M_\xi + 2\pi$, and $\zeta\to \zeta - 2\pi \Gamma$. 
%As clearly shown in the figure, these transformations are equivalent to 
%(\ref{discrete_shift}),  $\theta^i \to \theta^i + 2\pi n^i$, where $n^i$ are integers.
}
\label{fig:discrete}
\end{center}
\end{figure}

Using (\ref{zeta_xi}), the original field variable $\theta^i$ can be expressed in terms of 
$\zeta$ and $\xi$ as
%the canonically normalized gauge-invariant axion
%$\xi$ and the gauge-variant $\zeta$ providing the longitudinal component of the $U(1)_A$ gauge boson:
\bea
\label{theta_paramet}
\theta^i =k^i{\zeta}+
\frac{\sum_j(G^{-1})^{ij}\tilde k_j}{\sum_{ij}(G^{-1})^{ij}\tilde k_i\tilde k_j}\,\frac{\xi}{M_\xi}.
\eea
To identify the periodicities of $\xi$ and $\zeta$, let us consider how $\xi$ and $\zeta$
transform under
\bea
\label{discrete_shift}
\theta^i \to \theta^i+2\pi n^i\eea
for generic integers $n^i$, which correspond to the discrete gauge transformations generated by
(\ref{shift_gauge}). 
With the identity (\ref{complete}),  one can make the decomposition
\bea
\label{decom}
\frac{\sum_j (G^{-1})^{ij}\tilde k_j}{\sum_{i,j} (G^{-1})^{ij}\tilde k_i\tilde k_j}
% =\frac{\sum_{p,j}\delta^i_p(G^{-1})^{pj}\tilde k_j}{\sum_{i,j} (G^{-1})^{ij}\tilde k_i\tilde k_j}
%=\sum_{pj}(G^{-1})^{pj}\left(k^i\tilde\ell_p+\ell^i\tilde k_p\right)\tilde k_j
=\ell^i+ \Gamma(G,\vec k) k^i,
\eea
where\bea\label{Gamma}
\Gamma(G,\vec k) =\frac{1}{{\rm gcd}(\vec k)}\,\frac{\sum_{ij}(G^{-1})^{ij}\tilde \ell_i\tilde k_j}{\sum_{ij}(G^{-1})^{ij}\tilde k_i\tilde k_j},\eea
and  rewrite (\ref{theta_paramet}) as
\bea
\label{theta_exp}
\theta^i = k^i \zeta  + 
\Big(\ell^i + \Gamma(G, \vec k) k^i\Big) \frac{\xi}{M_\xi}.\eea
Note that if one chooses different solutions of (\ref{vec_dual}), e.g. $\vec\ell^\prime$ and $\vec{\tilde \ell}^\prime$ in
(\ref{degenerate}), the corresponding $\Gamma$ is shifted as
\bea
\label{degenerate2}
\Gamma(G,\vec k)\,\,\rightarrow\,\, \Gamma^\prime(G,\vec k)=\Gamma(G,\vec k)-\frac{q}{{\rm gcd}(\vec k)}.\eea
It is now straightforward to see that the discrete  transformation (\ref{discrete_shift}) results in
\bea
\zeta  &\to& \, \zeta+ \frac{2\pi}{{\rm gcd}(\vec k)}\sum_i\tilde\ell_i n^i -2\pi \Gamma(G,\vec k)\sum_i \tilde k_i n^i, \nonumber\\
\xi   &\to&\,  \xi  +2\pi M_\xi \sum_i\tilde k_i n^i,\eea
which are generated by 
\bea
\label{Z_xi1}
 &&\mathbb{Z}_\zeta:\,\,\, \zeta \to\, \zeta +\frac{2\pi}{{\rm gcd}(\vec k)}, \quad \xi \to\, \xi,\nonumber \\
 && \mathbb{Z}_\xi:\,\,\, \xi \to \, \xi +2\pi M_\xi(G,\vec k), \quad \zeta \to \, \zeta - 2\pi \Gamma(G, \vec k).\eea
In Fig.\ref{fig:discrete}, we depict $\mathbb{Z}_\zeta$ and $\mathbb{Z}_\xi$ in the axion moduli space of  $\theta^i$.
The discrete symmetry $\mathbb{Z}_\zeta$ involves only a shift of $\zeta$, so $\zeta$ is by itself a periodic field with the field range
\bea 
\Delta \zeta= \frac{2\pi}{{\rm gcd}(\vec k)}.\eea
 On the other hand, both $\xi$ and $\zeta$ are shifted under $\mathbb{Z}_\xi$, so $\xi$ is periodic with the field range \bea
 \Delta\xi=2\pi M_\xi(G,\vec k)\eea 
{\it only when} the accompanying shift $\zeta\to \,\zeta-2\pi \Gamma(G,\vec k)$ is taken into account. 
Note that $\Delta\zeta$ corresponds to
the volume (length) of the $U(1)_A$ gauge orbit in the axion moduli space and the coordinate direction $\partial/\partial\xi$
is normal to the $U(1)_A$ gauge orbit w.r.t the metric $G_{ij}$ (See Fig.~\ref{fig:discrete}), so that
\bea
M_A \Delta\zeta \Delta\xi={\rm Vol}(\vec\theta)=(2\pi)^2 \sqrt{{\rm  det}(G_{ij})},
\eea where
${\rm Vol}(\vec \theta)$ is the volume of the full axion moduli space.

In the original description using $\theta^i$, all non-derivative couplings of $\theta^i$ are  quantized  to be manifestly invariant under the $2\pi$ shifts of $\theta^i$.  However, in the description using $\xi$ and $\zeta$, which is more convenient for describing low energy physics below the St\"uckelberg mass $M_A$, the connection between the periodicity  and the non-derivative couplings of $\xi$ is less transparent as
one needs to  include the consequences of the accompanying shift $\zeta\to \,\zeta-2\pi \Gamma(G,\vec k)$.
This calls for a care when one attempts to deduce the field range $\Delta\xi$ from the couplings such as
$\xi\,G^a_{\mu\nu}\tilde G^{a\mu\nu}$ \cite{Shiu:2015uva,Fonseca:2019aux}.   In fact, the discrete symmetries $\mathbb{Z}_\zeta$ and $\mathbb{Z}_\xi$
have different realizations which are applicable even in the low energy limit where the massive $A_\mu-\partial_\mu\zeta$ is integrated out.  Combining them
with the $U(1)_A$ transformations (\ref{u1a}) for 
$\Lambda=-2\pi/{\rm gcd}(\vec k)$ (for $\mathbb{Z}_\zeta$) 
and $\Lambda=2\pi \Gamma(G,\vec k)$ (for $\mathbb{Z}_\xi$), one finds the  
equivalent discrete symmetries under which only the light fields transform:
\bea
\label{Z_xi2}
 && \mathbb{Z}_\zeta^\prime:
 \quad \xi \to\, \xi,\quad \Phi \to\, e^{i 2\pi q_\Phi/{\rm gcd}(\vec k)}\Phi, \nonumber \\
&&\mathbb{Z}_\xi^\prime : \quad \xi \to \, \xi +2\pi M_\xi, \quad \Phi \to\, e^{-i2\pi q_\Phi \Gamma}\Phi,
\eea 
where $\Phi$ denotes the generic  matter fields in the model.  Obviously 
 $\mathbb{Z}_\zeta^\prime$ corresponds to the discrete subgroup  of $U(1)_A$ unbroken by the  St\"uckelberg mechanism.  
As for  $\mathbb{Z}_\xi^\prime$ which is associated with the periodicity of $\xi$,
% involves non-trivial transformation of $U(1)_A$-charged $\Phi$.
% so the connection between $\Delta\xi$ and $\xi G^{a\mu\nu}\tilde G^a_{\mu\nu}$ is not manifest yet. 
one can  make 
the following $\xi$-dependent field redefinition:
\bea
\label{redef}
\Phi \,\,\to\,\,\, \exp\left( -i  q_\Phi \Gamma(G,\vec k)\frac{\xi}{M_\xi}\right)\Phi,\eea
after which the redefined $\Phi$ does not transform anymore. Then  
the discrete gauge symmetry for the periodicity of $\xi$ involves {\it only} a shift of $\xi$:
\bea
\label{Z_xi3}
\mathbb{Z}_\xi^{\prime\prime}: \quad \xi \to\, \xi +2\pi M_\xi.\eea
As it should be, the field redefinition (\ref{redef}) modifies the 
non-derivative couplings of $\xi$ in such a way that in the new field basis all non-derivative couplings are integer-multiples of $1/M_\xi$, so manifestly consistent with the axion periodicity
$\xi\equiv \xi+2\pi M_\xi$. It modifies also the derivative couplings of $\xi$ by generating 
%Note that the lagrangian in the new field basis is modified the couplings of $\xi$ in the new field basis.
\bea
\label{derivative}
\Delta {\cal L}_{\rm derivative} =  \frac{\Gamma(G,\vec k)}{M_\xi}
 \partial_\mu \xi(x) \sum_\Phi  J^\mu_\Phi,   
%+2\pi\Gamma\left(\sum_I q_I {\rm Tr}(T^2_a(\psi_I)) \right) \frac{\xi(x)}{\Delta\xi}G^{a\mu\nu}\tilde G^a_{\mu\nu}+...,  
\eea
where $J^\mu_\Phi$ is the $U(1)_A$ current of the matter field $\Phi$.

Let us see the connection between the axion periodicity and the axion couplings to matter and gauge fields more explicitly.
We first consider the coupling to $SU(N_c)$ gauge fields. 
Using (\ref{theta_exp}), the couplings in the original field basis can be decomposed 
as
% as in \cite{Shiu,Fonseca}.
\bea
\label{axion_instanton}
&&\frac{1}{32\pi^2}\Big(\sum_i r_i\theta^i \Big) G^a_{\mu\nu}\tilde G^{a\mu\nu}=  \frac{1}{32\pi^2}\Big(\Big(\sum_ir_ik^i\Big)\zeta+\frac{\xi}{f_\xi}\Big) G^a_{\mu\nu}\tilde G^{a\mu\nu}, \eea
where   
\bea
\label{axion_scales}
%\frac{1}{f_\zeta}&=& \frac{\sum_i r_ik^i}{M_A}=\frac{\sum_i r_ik^i}{{\rm gcd}(k^1,k^2)} \,\frac{2\pi}{\Delta\zeta}
%\nonumber \\
 \frac{1}{f_\xi} &=&  \frac{1}{M_\xi} \Big(\sum_i r_i \ell^i  +
%\left(\frac{\sum_{ij}(G^{-1})^{ij}\tilde \ell_i\tilde k_j}{\sum_{ij}(G^{-1})^{ij}\tilde k_i\tilde k_j}\right)
\Gamma(G,\vec k)\sum_i r_i k^i\Big) .
 \eea
%with
%\bea
%R(G,\vec k, \vec r)=
%\frac{\sum_{ij} (G^{-1})^{ij} r_i\tilde k_j}{\sum_{ij}(G^{-1})^{ij}\tilde k_i\tilde k_j}.
% \quad \Delta \xi=2\pi M_\xi.
%\eea 
As we have anticipated, the coupling of $\zeta$ is manifestly consistent with the periodicity
$\zeta\equiv \zeta + 2\pi/{\rm gcd}(\vec k)$. On the other hand,
the coupling of $\xi$, i.e. $1/f_\xi$, 
%the first piece in the bracket for the coupling $1/f_\xi$ has an integer value, the second piece has
contains a $G_{ij}$-dependent continuous piece in the unit of $1/M_\xi$,
%  determined by $\Gamma(G,\vec k)$, 
%a $G_{ij}$-dependent {\it continuous} value in the unit of $1/M_\xi$,
and therefore is {\it not} manifestly consistent with the periodicity
$\xi\equiv \xi+2\pi M_\xi$. This is not surprising  since we don't include yet the effect of the discrete shift of $\zeta$ in $\mathbb{Z}_\xi$, or
of the phase rotation of $\Phi$ in $\mathbb{Z}^\prime_\xi$, or of the field redefinition (\ref{redef}) for $\mathbb{Z}_\xi^{\prime\prime}$. 
As a specific choice,  let us make the field redefinition (\ref{redef}).  One of its consequences is   
%\bea
%\label{redef1}
%\Phi \,\,\to\,\,\, e^{i  q_\Phi \Gamma \xi/M_\xi}\Phi,\eea
the following change of lagrangian density
through the anomalous variation  of the path integral measure of $\psi_I$:
%This should be modified by the field redefinition (\ref{redef}) to a value being an integer-multiple of $2\pi/\Delta$ as 
%(\ref{redef}) 
% which is associated with 
%the mixed $U(1)_A\times [SU(N_c)]^2$ anomaly: 
\bea
\Delta {\cal L}_{\rm anomaly}&=&\frac{1}{32\pi^2}\frac{\Gamma(G,\vec k)}{M_\xi}\Big(2\sum_I q_I {\rm Tr}(T^2_a(\psi_I)) \Big) \xi\, G^{a\mu\nu}\tilde G^a_{\mu\nu}\nonumber \\
&=& -\frac{1}{32\pi^2}\frac{\Gamma (G,\vec k)}{M_\xi}
\Big(\sum_i r_i k^i\Big) \xi\, G^{a\mu\nu}\tilde G^a_{\mu\nu}, \eea
where we used the anomaly cancellation condition (\ref{gs_condition}) for the latter expression. 
Including this change,  the continuous piece of $1/f_\xi$ is cancelled and the coupling is modified  to a quantized value which is manifestly consistent with
the periodicity $\xi\equiv \xi+2\pi M_\xi$: 
\bea
\frac{1}{f_\xi} \,\,\to\,\,\,   \frac{1}{M_\xi} \sum_i r_i\ell^i  .
\eea

The underlying $U(1)_A$ gauge symmetry assures that such modification applies for all non-derivative couplings of $\xi$. 
To see this, let us consider the coupling of $\xi$ to an operator ${\cal O}(\Phi)$ $({\cal O}(\Phi)=\psi_I\psi_J ,\,\phi_P\psi_I\psi_J, ...)$ whose $U(1)_A$ charge is $q_{\cal O}$, which would originate from
\bea
 \exp\Big( i\sum_i n^{\cal O}_i\theta^i\Big) {\cal O}(\Phi), \eea
 %\,=e^{2\pi i\left(\left(\sum_i n_i k^i\right)\frac{\zeta}{\Delta\zeta}+ 
%\left(\frac{\sum_{ij} (G^{-1})^{ij} r_i\tilde k_j}{\sum_{ij}(G^{-1})^{ij}\tilde k_i\tilde k_j}\right)\frac{\xi}{\Delta\xi}%\right)}\, {\cal O}(\Phi) ,\eea
where the $U(1)_A$ invariance requires that the integer-valued coefficients $n_i^{\cal O}$ satisfy
\bea
\sum_i n_i^{\cal O} k^i = q_{\cal O}.\eea
As in the case of the coupling to $SU(N_c)$ gauge fields, 
$\sum_i n_i^{\cal O} \theta^i$  can be expressed in terms of $\zeta$ and $\xi$ as
\bea
\sum_i n_i^{\cal O} \theta^i
%&=&\left(\sum_i n_i k^i\right)\zeta+ 
%\left(\frac{\sum_{ij} (G^{-1})^{ij} n_i\tilde k_j}{\sum_{ij}(G^{-1})^{ij}\tilde k_i\tilde k_j}\right)\frac{2\pi\xi}{\Delta\xi} %\nonumber \\
= \Big(\sum_i n_i^{\cal O} k^i\Big)\zeta +
\Big(\sum_i n_i^{\cal O} \ell^i + \Gamma(G,\vec k)\sum_i n_i^{\cal O} k^i\Big)\frac{\xi}{M_\xi}. 
\eea
%where we used the expression (\ref{theta_exp}).
Again, under the field redefinition (\ref{redef}), 
%\bea
%\label{redef2}
%\Phi \,\,\to\,\,\, e^{i q_\Phi \Gamma \xi/M_\xi}\Phi,\eea
 ${\cal O}(\Phi)$ transforms as
%together with the $U(1)_A$ invariance condition (\ref{u1_condition}),
\bea
{\cal O}(\Phi) \, \, 
  \rightarrow \, \, 
\exp \left(-i q_{\cal O} \Gamma(G, \vec k) \frac{\xi}{M_\xi}\right){\cal O}(\Phi),
\eea
%\bea
%\psi_I\psi_J &\rightarrow & e^{-i2\pi (q_I+q_J)\Gamma\xi(x)/\Delta\xi}\psi_I\psi_J,\nonumber \\
%\phi_P\psi_I\psi_J &\rightarrow & e^{-i2\pi  (q_P+q_I+q_J)\Gamma\xi(x)/\Delta\xi}\phi_P\psi_I\psi_J 
%\eea
which results in the quantized non-derivative couplings of $\xi$ as
%to the values manifestly consistent with 
%the axion periodicity $\xi\equiv \xi+\Delta\xi$:
\bea
\exp\Big( i\sum_i n_i^{\cal O}\theta^i \Big) {\cal O}(\Phi)  \, \, 
  \rightarrow \, \,  \exp\Big( i \sum_i n_i^{\cal O}\ell^i \frac{\xi}{M_\xi}\Big) {\cal O}(\Phi).\eea
So, in the new field basis after the field redefinition (\ref{redef}), all non-derivative couplings of $\xi$ are given by inter-multiples of $1/M_\xi$, and therefore manifestly consistent with the axion periodicity $\xi\equiv\xi+2\pi M_\xi$, while the derivative couplings are shifted by the additional terms in (\ref{derivative}).

Let us summarize the above discussions with a low energy effective theory obtained by integrating out
the massive $U(1)_A$ gauge boson.
% and derive  the low energy effective lagrangian  of the  gauge-invariant light axion combination $\xi$. 
The $U(1)_A$ gauge invariance admits to choose the unitary gauge $\zeta=0$ and integrate out $A_\mu$ using its equation of motion. For the model of (\ref{model1}), this results in the effective lagrangian density
of the light axion $\xi$, gauge fields $G^a_\mu$ and matter fields $\Phi=(\phi_P, \psi_I)$, which is given by
\bea
\label{eff1}
{\cal L}_I &=& \frac{1}{2} \partial_\mu\xi\partial^\mu\xi
-\frac{1}{4g_a^2} G^a_{\mu\nu} G^{a\mu\nu} + \frac{1}{32\pi^2}\sum_i r_i (\ell^i + \Gamma (G,\vec k) k^i )\frac{\xi}{M_\xi} G^a_{\mu\nu}\tilde G^{a\mu\nu} \nonumber \\
&&+\, 
\sum_P |D_\mu\phi_P|^2 + 
\sum_I\bar\psi_I  i  \bar\sigma^\mu D_\mu  \psi_I 
 -\left(\mu_{IJ}e^{i\sum_i n^{IJ}_ i (\ell^i+ \Gamma(G,\vec k)k^i)\xi/M_\xi}\psi_I\psi_J+{\rm h.c.}\right)\nonumber \\
 &&
 -\, \left(\lambda_{IJP} e^{i\sum_i n^{IJP}_i (\ell^i+ \Gamma(G,\vec k) k^i) \xi/M_\xi }\phi_P\psi_I\psi_J+{\rm h.c.}\right)\nonumber \\
 &&+\,\frac{\partial_\mu\xi}{M_\xi}\sum_i  (\ell^i+ \Gamma(G,\vec k) k^i)J^\mu_i+{\cal O}\left(\frac{1}{M_A^2}\right),
\eea 
where ${\cal O}(1/M_A^2)$  stands for  the higher-dimensional effective interactions generated by the exchange of the massive
$U(1)_A$ gauge field, and the ellipsis denotes the other possible terms including the potential of $\xi$ and $\phi_P$. Here we are using the same matter field basis as in the original model and
the periodicity of $\xi$ is ensured by 
the discrete gauge symmetry (\ref{Z_xi2}):
 \bea
\mathbb{Z}_\xi^\prime: \quad \xi \to \, \xi +2\pi M_\xi, \quad \Phi \to\, e^{-i2\pi q_\Phi \Gamma}\Phi,
\eea 
where $M_\xi$, $\Gamma(G, \vec k)$  and  $\vec\ell, \vec {\tilde k}, \vec{\tilde\ell}$ are 
defined in (\ref{M_xi}), (\ref{Gamma}) and (\ref{vec_dual}), respectively. 
%\bea
%M^2_\xi =\frac{1}{\sum_{ij}(G^{-1})^{ij}\tilde k_i \tilde k_j}, \quad
%\Gamma(G,\vec k) =\frac{1}{{\rm gcd}(\vec k)}\,\frac{\sum_{ij}(G^{-1})^{ij}\tilde \ell_i\tilde k_j}%{\sum_{ij}(G^{-1})^{ij}\tilde k_i\tilde k_j},
%\eea
%and the integer-valued $\vec{\tilde k}, \vec{\ell}, \vec{\tilde \ell}$
%are determined by 
%\bea
%\label{constraint}
%\vec k \cdot \vec{\tilde k} =0,  \quad \vec\ell\cdot \vec{\tilde \ell} =0, \quad \vec %\ell\cdot\vec{\tilde k}=\frac{\vec k\cdot\vec{\tilde \ell}}{{\rm gcd}(\vec k)}=1,\quad {\rm gcd}(\vec %{\tilde k})=1,\eea
%where  $\vec k=(k^1, k^2)$ is the $U(1)_A$ charge vector of the original axion field $%\vec\theta=(\theta^1,\theta^2)$.
In the above, all non-derivative axion couplings are decomposed into two pieces, a piece quantized in the unit of $1/M_\xi$ and the other continuous piece proportional to $\Gamma(G,\vec k)$.  Obviously such decomposition is not unique, but has an ambiguity parametrized by integer as (\ref{degenerate}) and (\ref{degenerate2}). 
%Indeed $\vec \ell,\vec{\tilde \ell}$ satisfying (\ref{constraint}) and the corresponding $\Gamma$  have degenerate solutions parametrized by an arbitrary integer $n$:
%\bea
%\vec\ell+n\frac{\vec k}{{\rm gcd}(\vec k)}, \quad \vec{\tilde\ell}-n\vec{\tilde k},\quad \Gamma -\frac{n}{{\rm gcd}(\vec k)},\eea
%which corresponds to the integer ambiguity in the decomposition of couplings into the quantized  and continuous pieces.
 The axion couplings in (\ref{eff1}) assures that the continuous parts of all non-derivative couplings of $\xi$  can be rotated away  by
the field redefinition 
\bea
\Phi \,\,\to\,\,\, e^{-i  q_\Phi \Gamma \xi/M_\xi}\Phi,\eea
after which the effective lagrangian density takes the form
\bea
\label{eff2}
{\cal L}_{II} &=& \frac{1}{2} \partial_\mu\xi\partial^\mu\xi
-\frac{1}{4g_a^2} G^a_{\mu\nu} G^{a\mu\nu} +
 \frac{1}{32\pi^2} \Big(\sum_i r_i \ell^i \Big) \frac{\xi}{M_\xi} G^a_{\mu\nu}\tilde G^{a\mu\nu} \nonumber \\
&&+\, 
\sum_P D_\mu\phi_P^* D^\mu \phi_P+ 
\sum_I\bar\psi_I  i  \bar\sigma^\mu D_\mu  \psi_I 
 -\left(\mu_{IJ}e^{i (\sum_i n^{IJ}_ i\ell^i) \xi/M_\xi}\psi_I\psi_J+{\rm h.c}\right)\nonumber \\
 &&
 -\, \left(\lambda_{IJP} e^{i (\sum_i n^{IJP}_i\ell^i) \xi/M_\xi }\phi_P\psi_I\psi_J+{\rm h.c.}\right)\nonumber \\
 &&+\,\frac{\partial_\mu\xi}{M_\xi}\Big(\sum_i  (\ell^i+\Gamma(G,\vec k) k^i )J^\mu_i +\Gamma(G,\vec k)\sum_\Phi J^\mu_\Phi\Big)
+{\cal O}\left(\frac{1}{M_A^2}\right) ,
\eea 
so all non-derivative couplings of $\xi$ are quantized to be manifestly consistent with the axion periodicity
$\xi\equiv \xi+2\pi M_\xi$.
%\bea
%Z_\xi^{\prime\prime}: \quad \xi \to \, \xi +2\pi M_\xi, \quad \Phi \to \, \Phi.
%\eea 

In the above, we presented the low energy effective theory of the model (\ref{model1}) in two different field basis. 
%In the second field basis yielding the effective lagrangian ${\cal L}_{II}$, all non-derivative couplings of $\xi$ are %manifestly consistent with the axion periodicity $\xi\equiv \xi+2\pi M_\xi$, which might be useful for deducing  the axion field range from the  couplings. However 
It should be stressed that  axion couplings to matter and/or gauge fields are basis-dependent, e.g. vary under axion-dependent phase rotation of matter fields,
%even when the axion, gauge fields and matter fields are all canonically normalized, 
while their physical consequences should be basis-independent.
In the next section, we will discuss this issue with a simple example.
%the results are independent of the choice of field basis as will be shown explicitly 
% in the next section with simple example.
% Of course, one can compute various basis-independent (reparameterization-invariant) physical quantities such as 
%the effective potential and 1PI amplitudes of $\xi$  from those basis-dependent couplings.

\subsection{Generalization to multiple ($N>2$) axions} \label{Large_N}
 
%So far, we have limited our discussion to the case of two axions which have a St\"uckelberg mixing with single $U(1)_A$ gauge %boson. 
It is in fact straightforward to generalize the discussion to more general cases, for instance models with $N(>2)$ axions having the St\"uckelberg mixings with $(N-1)$  $U(1)$ gauge bosons.  In such models, the gauge invariant kinetic terms of axions can be written as
\bea \label{kinetic_multi}
{\cal L}_{\rm kin}=
%-\sum_{\alpha=1}^{N-1}\frac{1}{4 g_\alpha^2}F_{\mu\nu}^\alpha F^{\alpha \mu\nu} + 
\frac{1}{2} \sum_{ij=1}^{N} G_{ij} \Big(\partial_\mu\theta^i - \sum_{\alpha=1}^{N-1} k^i_\alpha A^\alpha_\mu \Big) \Big(\partial^\mu \theta^j 
- \sum_{\beta=1}^{N-1} k^j_\beta A^{\beta\mu}\Big) 
%- \frac{1}{4}\sum_{\alpha\beta} f_{\alpha\beta} F^\alpha_{\mu\nu} F^{\beta\mu\nu},
\eea 
and the $U(1)$ gauge transformations of the fields are given by 
\bea \label{multi_U1}
U(1)_\alpha:\quad  A^\alpha_\mu \to A^\alpha_\mu  + \partial_\mu \Lambda^\alpha, 
 \quad \theta^i \to \theta^i  +  \sum_{\alpha}k^i_\alpha \Lambda^\alpha,\quad 
\Phi \to  e^{ - i \sum_\alpha q_{\Phi\alpha} \Lambda^\alpha} \Phi, 
\eea 
where $\Phi$ denotes the gauge-charged matter fields in the model. Again $\theta^i$ and $\Lambda^\alpha$ are normalized to  have the $2\pi$ periodicity, and then all $U(1)$ charges $k^i_\alpha$ and $q_{\Phi\alpha}$ have integer values.
Constructing a complete set of the integer-valued vectors and dual vectors in the $\theta$-space is also useful here. 
For $N-1$ linearly independent vectors, 
\bea 
\vec k_{r \alpha}
= (k_{r\alpha}^1,\cdots, k^N_{r\alpha})
\equiv    \frac{(k^1_\alpha, \cdots, k^N_\alpha)}{{\rm gcd}(\vec k_\alpha)} 
\quad {\rm for}\quad \alpha = 1,2,\cdots, N-1, 
\eea 
we can find the remaining vector $\vec\ell$ and the
$N$   dual vectors $\vec{\tilde k}, \vec{\tilde \ell}^\alpha$ 
from
\bea \label{vec_dual_multi}
\vec {\tilde k}\cdot \vec k_{r\alpha}  =0,\quad 
\vec{\tilde k}\cdot \vec \ell = 1,\quad 
\vec{\tilde \ell}^\alpha \cdot \vec k_{r\beta} = \delta^\alpha_\beta,\quad 
\vec{\tilde \ell}^\alpha\cdot \vec \ell = 0 \quad {\rm for}\quad \alpha, \beta = 
1,2,\cdots, N-1.
\eea 
As in the case of two axion model, 
$\vec{\tilde k}$ is  determined uniquely (up to sign) by the $U(1)_A$ charge vectors $\vec k_{r\alpha}$  as 
\bea \label{ktilde_multi}
\tilde k_i =\pm \det\left[\begin{array}{cccc} \delta^1_i  & \delta^2_i &\cdots & \delta^N_i  \\ 
 k^1_{r1} &   k^2_{r1} & \cdots &   k^N_{r1} \\ 
  k^1_{r2} &   k^2_{r2} &\cdots & k^N_{r2} \\  
\vdots   & \vdots & \ddots & \vdots    \\ 
  k^1_{r N-1} &  k^2_{r N-1} &\cdots &   k^N_{r N-1}  \end{array}\right].
\eea  
For given $\vec k_{r\alpha}$ and $\vec{\tilde k}$, 
the corresponding $\vec \ell$ and $\vec{\tilde \ell}^\alpha$ are not unique as the conditions in (\ref{vec_dual_multi}) are 
invariant under the reparameterization
\bea 
\label{reparameterization}
\vec\ell\,\, \to \,\,
\vec\ell^\prime =\vec \ell +\sum_\alpha q^\alpha \vec k_{r\alpha}, \quad
\vec{\tilde \ell}^\alpha \,\,\to \,\, \vec{\tilde\ell}^{\alpha\prime}= \vec{\tilde \ell}^\alpha -q^\alpha \vec{\tilde k},\eea
where   $q^\alpha$ ($\alpha=1,2,\cdots, N-1$) are arbitrary independent integers.
%\quad (q^\alpha\in \mathbb{Z}).
However such degeneracy of $\vec\ell$ and $\vec{\tilde\ell}^\alpha$ does not matter to us as all solutions 
%set of the integer-valued $\vec k_{r\alpha}, \vec\ell, \vec{\tilde k}, \vec{\tilde\ell}_\alpha$ satisfying   
satisfy the common completeness relation
\bea  \label{complete2}
\sum_\alpha k^i_{r\alpha} \tilde \ell^\alpha_j +  \ell^i \tilde k_j  =\delta^i_j. 
\eea

%Because $N-1$ axions are eaten by the  gauge bosons through the St\"uckelberg mechanism, 
Again, one can express the original axion fields $\theta^i$ in terms of  the canonically normalized gauge-invariant $\xi$ and the gauge-variant $\zeta^\alpha$ eaten  by $A_\mu^\alpha$:
\bea \label{theta_decom}
\theta^i  =   \sum_{\alpha=1}^{N-1} k^i_\alpha \zeta^\alpha + \frac{\sum_{j=1}^N (G^{-1})^{ij}\tilde k_j}{\sum_{ij} (G^{-1})^{ij}\tilde k_i \tilde k_j } \frac{\xi}{M_\xi} \eea
for which
the axion kinetic terms (\ref{kinetic_multi}) take the familiar form
\bea
{\cal L}_{\rm kin} =
%-\sum_{\alpha=1}^{N-1}\frac{1}{4 g_\alpha^2}F_{\mu\nu}^\alpha F^{\alpha \mu\nu} + 
\frac{1}{2}(\partial^\mu\xi)^2 +\frac{1}{2}\sum_{\alpha\beta} (M_A^2)_{\alpha\beta}(A_\mu^\alpha-\partial_\mu\zeta^\alpha)(A^{\mu \beta}-\partial^\mu\zeta^\beta),\eea
where 
% the mass matrix of the $U(1)$ gauge bosons is   is determined  by  matrix 
\bea 
 (M_A^2)_{\alpha\beta} =  
 \sum_{ij} G_{ij} k^i_\alpha k^j_\beta, \quad M^2_\xi =\frac{1}{\sum_{ij} (G^{-1})^{ij} \tilde k_i \tilde k_j}.
\eea
Obviously  $(M_A^2)_{\alpha\beta}$ is the mass matrix of the $(N-1)$ massive $U(1)$ gauge bosons. As in the case of two axion model, we will see that $M_\xi$ corresponds to the decay constant of $\xi$, i.e. the field range of $\xi$ is given by  $\Delta\xi=2\pi M_\xi$.  
From (\ref{theta_decom}), we find also 
%$\xi$ and $\zeta^\alpha$ also can be represented by the original axion fields as
 \bea 
\frac{\xi}{M_\xi} =   \sum_i \tilde k_i \theta^i,  
\quad
\zeta^\alpha = \sum_{ij\beta}\left(M^{-2}_A\right)^{\alpha\beta} G_{ij} k^i_\beta \theta^j, 
\eea

Using the identity (\ref{complete2}), one can rewrite (\ref{theta_decom}) as
\bea
\theta^i = \sum_{\alpha=1}^{N-1}  k^i_\alpha  \zeta^\alpha 
+ \Big(\ell^i + \sum_{\alpha=1}^{N-1}  \Gamma^\alpha(G, \vec k_\alpha) k^i_\alpha\Big)  \frac{\xi}{M_\xi},
\eea 
where
% the kinetic metric dependent continuous parameter $\Gamma^\alpha$ is defined by 
\bea 
\Gamma^\alpha(G, \vec k_\alpha) =  \frac{1}{{\rm gcd}(\vec k_\alpha)}\frac{\sum_{ij} (G^{-1})^{ij} \tilde\ell_i^\alpha \tilde k_j}{\sum_{ij} (G^{-1})^{ij} \tilde k_i \tilde k_j}.
\eea 
With the above expression,
one can see that the discrete gauge symmetries for the $2\pi$ periodicities of $\theta^i$, i.e. 
\bea
 \mathbb{Z}_i:\,\,\, \theta^i \,\,\to\, \theta^i+2\pi\quad (i=1,\cdots, N),\eea
are generated by
\bea
\mathbb{Z}_{\zeta^\alpha}:&& \zeta^\alpha \,\,\to\, \zeta^\alpha+\frac{2\pi}{{\rm gcd}(\vec k_\alpha)}\quad  (\alpha=1,\cdots, N-1), \nonumber \\
  \mathbb{Z}_\xi: &&  \xi \,\,\to\, \xi +2\pi M_\xi, \quad \zeta^\alpha \,\,\to\, \zeta^\alpha -2\pi \Gamma^\alpha(G,\vec k_\alpha). \eea
One can consider also the equivalent discrete symmetries which do not involve a transformation of the massive $\zeta^\alpha$: 
\bea
\mathbb{Z}_{\zeta^\alpha}^\prime &=& \mathbb{Z}_{\zeta^\alpha}\times \left. U(1)_\alpha\right|_{\Lambda^\alpha = \frac{-2\pi}{{\rm gcd}(\vec k_\alpha)}}: 
\quad \xi \to \xi ,\quad \Phi \to e^{  i 2\pi q_{\Phi_\alpha} /{\rm gcd}(\vec k_\alpha)} \Phi,\nonumber\\
\mathbb{Z}_\xi^{\prime} &=& \mathbb{Z}_\xi\times \prod_\alpha\left.U(1)_\alpha\right|_{\Lambda^\alpha=2\pi \Gamma^\alpha}: \quad \xi \to \, \xi +2\pi M_\xi, \quad \Phi \to\, e^{- i 2\pi \sum_\alpha q_{\Phi_\alpha} \Gamma^\alpha}\Phi.
\eea

As for $\mathbb{Z}_\xi^{\prime}$ which is associated with the periodicity of $\xi$,  
one can  make the $\xi$-dependent field redefinition 
\bea \label{field_redef}
\Phi \,\,\to\,\,\, 
\exp\left( -i  \sum_\alpha q_{\Phi_\alpha} \Gamma^\alpha(G, \vec k_\alpha)   \frac{\xi}{M_\xi} \right) \Phi ,\eea
after which the periodicity of $\xi$ is assured by
\bea
\mathbb{Z}_\xi^{\prime\prime}: \quad \xi \to \, \xi +2\pi M_\xi, \quad \Phi \to \, \Phi.
\eea 
As in the case of two axion model, the above field redefinition  provides a field basis for which
 all non-derivative couplings of $\xi$ are quantized in the unit of $1/M_\xi$ in a manner manifestly consistent with the axion periodicity  $\xi\equiv \xi+2\pi M_\xi$.

Models of $N(>2)$ axions which have the St\"uckelberg mixings with $(N-1)$ $U(1)$ gauge bosons exhibit several distinctive features in the  limit $N\gg 1$. First of all, in such limit $\xi$ has a  field range {\it exponentially suppressed} relative to the original axion scales encoded in the axion kinetic metric $G_{ij}$.
% of the angular axion fields $\theta^i$ with the $2\pi$ periodicity.  
%Let us remark few features of $M_\xi$ (\ref{Mxi}). 
To see this, let us  note that $M_\xi$ is bounded as
\bea 
%\frac{1}{ \tilde{\norm{k}}} \frac{1}{\sqrt{{\rm Tr}(G^{-1})}} < 
\frac{f_{\rm min}}{||\tilde{k}||}
 \leq  M_\xi \leq \frac{f_{\rm max}}{||\tilde{k}||},
%<\frac{1}{\tilde{\norm{k}}}  \sqrt{{\rm Tr}(G)}, 
\eea 
where $f^2_{\rm max}$ and $f^2_{\rm min}$ denote the maximum and minimum eigenvalues of $G_{ij}$, and
\bea
||\tilde k|| \equiv  \Big(\sum_i\tilde k_i \tilde k_i\Big)^{1/2}.\eea
%It is noted that the overall scale of $M_\xi$ is strictly related with the norm of the integer-valued vector $\vec{\tilde k}$, 
%$\tilde{\norm{k}} \equiv \sqrt{\sum_i\tilde k_i \tilde k_i}$.  
It was shown in \cite{Choi:2014rja} that $||\tilde{k}||$ determined by 
(\ref{ktilde_multi})
   grows exponentially in the limit $N\gg 1$:
\bea 
\label{large_k}
||\tilde{k}||
\sim  (k^i_{r\alpha})_{\rm rms}^{N-1}\sqrt{ N!},
\eea where 
$(k^i_{r\alpha})_{\rm rms}$ is the root-mean-square of the normalized $U(1)$ charges:
\bea
(k^i_{r\alpha})_{\rm rms} = \sqrt{\frac{\sum_{i\alpha} (k^i_{r\alpha})^2}{N(N-1)}}\quad
\Big(k^i_{r\alpha}=\frac{k^i_\alpha}{{\rm gcd}(\vec k_\alpha)}\Big).\eea
 In the clockwork axion models \cite{Choi:2014rja,Choi:2015fiu,
Kaplan:2015fuy, Giudice:2016yja}, a large value of
$||\tilde{k}||$  results in  an enlarged field range of the light axion combination as $||\tilde{k}||$ can be interpreted as the number of windings along the light axion direction.
On the other hand, 
in the St\"uckelberg axion models under discussion, a large $||\tilde{k}||$ means an enlarged volume of the gauge orbit of
$\prod U(1)_\alpha$ in the axion moduli space of $\theta^i$. As a consequence, it results in a reduction of the field range of the gauge-invariant axion combination which is normal to the gauge orbit of $\prod U(1)_\alpha$.
Specifically, for \bea
G_{ij}\sim f^2\delta_{ij},\eea the axion field range is reduced as 
%For a typical {\it naive} axion decay constant, $f$,  
%the true axion decay constant is hierarchically smaller than $f$ as 
\bea 
\label{m_xi_behavior}
 M_\xi \sim  \frac{1}{(k^i_{r\alpha})_{\rm rms}^{N-1}\sqrt{ N!}} \, f,
 \eea 
 which is exponentially smaller than the original axion scale $f$.

%which exponentially suppressed relative to the original axion scale $f$.
 
St\"uckelberg axion models in the limit $N\gg 1$  have an unusual feature which may cause a confusion in some case. 
In the original field basis without making any $\xi$-dependent field redefinition,
all  (both derivative and non-derivative) couplings of $\xi$ are determined simply by the couplings of $\theta^i$ and the wavefunction mixing between $\theta^i$ and $\xi$, which is given by (see(\ref{theta_decom}))
\bea
\label{wf_mixing}
\langle \theta^i|\xi\rangle =\frac{\sum_{j=1}^N (G^{-1})^{ij}\tilde k_j}{\sum_{ij} (G^{-1})^{ij}\tilde k_i \tilde k_j } \frac{1}{M_\xi}.\eea
As $\theta^i$ are angular fields with $2\pi$ periodicity, their couplings can be  described by 
dimensionless parameters, e.g. the integer coefficients $r_i, s_i, n_i^{IJ}, n_i^{IJP}$ for the non-derivative couplings in (\ref{model1}) and also the continuous parameters
$\kappa_{\psi_I}^i, \kappa_{\phi_P}^i$ describing the derivative couplings of the form:\bea
D_\mu\theta^i J^\mu_i = D_\mu\theta^i\left(\kappa_{\psi_I}^i\bar\psi_I\sigma^\mu\psi_I+i\kappa_{\phi_P}^i (\phi_P^*D_\mu\phi_P-\phi_P D_\mu\phi^*_P)+ ...\right).\eea
If the model does not involve any large dimensionless coupling or large number of fields, which might be required for a sensible UV behavior of the model,  those couplings of $\theta^i$ are all expected to be of order unity or smaller.  On the other hand, for $G_{ij}\sim f^2\delta_{ij}$, the wavefunction mixing (\ref{wf_mixing}) is bounded as
 \bea
\langle \theta^i|\xi\rangle 
%\frac{\sum_{j=1}^N (G^{-1})^{ij}\tilde k_j}{\sum_{ij} (G^{-1})^{ij}\tilde k_i \tilde k_j } \frac{1}{M_\xi}
\lesssim \frac{1}{||\tilde{k}||}\frac{1}{M_\xi} \sim \frac{1}{f}.
% \quad \mbox{for}\,\,\, G_{ij}\sim f^2\delta_{ij}, 
\eea
This implies that in the original field basis all (both derivative and non-derivative) couplings of $\xi$ are of the order of $1/f$ or  smaller if the couplings of $\theta^i$ are of order unity or smaller, which is exponentially weaker than the strength $\sim 1/M_\xi$  in the limit $N\gg 1$.
% which is naively expected from the field range. 
On the other hand, we already noticed that  after the $\xi$-dependent field redefinition (\ref{field_redef}), all non-derivative couplings of $\xi$  are quantized in the unit of $1/M_\xi$, so either exactly zero or of the order of $1/M_\xi$. If any of those quantized non-derivative couplings is nonzero, some derivative couplings should be of the order of  $1/M_\xi$ in the new field basis due to the pieces induced by the field redefinition (\ref{field_redef}).  
This means that   in the limit $N\gg 1$
 axion couplings to matter and gauge fields  
 have hierarchically different size  in the two field bases related by the field  redefinition (\ref{field_redef}).
Since all physical consequences of the model should be independent of the choice of field basis, if one uses the new field basis, then there should be a fine cancellation between the contributions from different couplings of ${\cal O}(1/M_\xi)$ to make the total result to be of the order of $1/f\ll 1/M_\xi$ as suggested by the couplings in the original field basis.

%Therefore, in the limit $N\gg 1$, axion couplings can have hierarchically  different size
%in the two different field bases related by the $\xi$-dependent field redefinition (\ref{field_redef}),  although both field %bases yield the same physical results. 
%We already noticed that after the $\xi$-dependent field redefinition (\ref{field_redef}), all non-derivative couplings of 
%$\xi$ are quantized in the unit of $1/M_\xi$, so are manifestly consistent with the axion periodicity
% $\xi\equiv \xi+2\pi M_\xi$. 
%However, to see how  the multiple $U(1)$ gauge symmetries protect $\xi$ from getting a mass, it is more convenient 
%to use the original field basis without making the field redefinition (\ref{field_redef}), for which

%that in models with the clockwork-type St\"uckelberg mixings between $N$ axions and $(N-1)$ $U(1)$ gauge bosons,
%$U(1)$ charges of the axions for the St\"uckelberg mixings with multiple $U(1)$ gauge bosons,
%there exists an accidental global  $U(1)_\xi$ symmetry involving a continuous shift of $\xi$, which is protected so well by 
% $\frac{1}{32\pi^2}\frac{\xi}{f_\xi} G^{a\mu\nu}\tilde G^a_{\mu\nu}$ should involve a large coefficient to be 
%compatible with 
%the underlying $U(1)$ gauge symmetries in the limit $N\gg 1$. As we will see, such a highly protected $U(1)_\xi$ has the same %origin as the exponential reduction of the axion field range $M_\xi$ by the St\"uckelberg mixings in the limit $N\gg 1$.

As a related feature, in St\"uckelberg axion models in the limit $N\gg 1$, $\xi$ is so well protected by the $(N-1)$ $U(1)$ gauge symmetries from getting a mass,
%the accidental global $U(1)_\xi$ symmetry involving the continuous shift
%\footnote{Generically the accidental symmetry $U(1)_\xi$  involves also model-dependent global phase rotations of 
%matter fields.}
%$\xi\rightarrow \xi+{\rm constant}$  
%is  well protected by the $(N-1)$ $U(1)$ gauge symmetries 
and therefore can be ultra-light in a natural way. This 
was noticed before in \cite{Bonnefoy:2018ibr} for a specific model with clockwork-type  $U(1)$ gauge charges of $\theta^i$.  
%i.e. $M_\xi\ll f$, 
In fact, this is a consequence of the exponentially large $||\tilde k||$, so a generic feature of  the St\"uckelberg axion models in the limit $N\gg 1$. 
To see this, let us consider the constraint on the axion potential from the $U(1)$ gauge symmetries. 
In the prescription where the $2\pi$ periodicities of all $\theta^i$ are manifest, any non-trivial axion potential should  be
 a periodic function of  the gauge-invariant combination of $\theta^i$, i.e.
% combination $\sum_i \tilde L_i\theta^i$, 
\bea
V_{\rm eff}= V_{\rm eff}\Big( \sum_i L_i\theta^i \Big)
%=V_{\rm eff}( \sum_i L_i\theta^i+2\pi), 
\eea
for integer coefficients 
% calculable integers determined by the integer-valued non-derivative couplings of $\theta^i$. On the other hand, 
%the invariance of the effective potential under
%$\prod_\alpha U(1)_\alpha$ requires that $L_i$ 
$L_i=L\tilde k_i$ where $L$ is a non-zero integer,
for which
\bea
\sum_i L_i\theta^i= L \frac{\xi}{M_\xi}.\eea
As $||\tilde{k}||$ is exponentially large  in the limit $N\gg 1$, some 
$L_i$ should be
exponentially large also. 
This means that any mechanism to generate an axion potential should provide  those large integer coefficient  $L_i$. 
The required  large $L_i$ might be achieved by introducing many degrees of freedom or operators with very high mass dimensions as  
discussed in \cite{Bonnefoy:2018ibr}. In any case, generically the requirement of exponentially large $L_i$  provides a strong constraint on the mechanism to generate an axion potential, and usually makes  the induced axion potential highly suppressed.
% as was extensively 
%from the non-derivative couplings of $\theta^i$ which are presumed to be of order unity, one needs an exponentially 
%many insertions of the couplings to generate an (either perturbative or non-perturbative) axion  potential, which would 
%generically result in a highly suppressed  $V_{\rm eff}$.
%As was pointed out in \cite{dudas},
% the suppression is so efficient that an 
%ultralight axion can be achieved with a not so large $N$.
In the next section, we will present an explicit model of  $N(>2)$ axions $\theta^i$ whose gauge charges for  $(N-1)$ $U(1)$ symmetries have a clockwork pattern \cite{Bonnefoy:2018ibr}, and
study the behavior of the model in the limit $N\gg 1$. 
% bosons axions, whose charges $k^i_\alpha$ are specially arranged. In this case, the anti-clockwork feature appears as 
%$M_\xi \sim q^{-N} f$  (roughly, it corresponds to $(k^i_{r\alpha})_{\rm rms} \sim q/\sqrt{N}$). 
%We will discuss further implication of this set-up in next section. 

\section{Implications with  examples}\label{sec:impl}

In the previous section, we derived the field range of the gauge-invariant axion combination $\xi$ and examined the structure of its couplings in generic models with the St\"uckelberg mixing between axions and $U(1)$ gauge bosons. In this section, we apply our results to the two specific examples to see some implications of our results explicitly. 

\subsection{An illustrative simple model}

% and compare them  
% We also identified the field basis for which all non-derivative couplings of
% $\xi$ are manifestly consistent with the axion periodicity 
%$\xi\equiv \xi+\Delta\xi$.
% with the related previous discussions. 
%In \cite{Shiu,Fonseca}, the axion field range $\Delta\xi$ was estimated through %the coupling (\ref{axion_instanton}) under the assumption that the corresponding %effective scale $2\pi f_\xi$ can be identified as  the field range of $\xi$.
% even in the case with $\sum_i r_ik_i\neq 0$, 
%which would result in  a variety of interesting possibilities.
% which were  extensively discussed in \cite{Shiu,Fonseca}. 
%One of the motivations of these works is to explore the possibility to
%generate $\Delta\xi$ much bigger than the original mass scales encoded in the %kinetic metric $G_{ij}$, e.g. the scales  $\sqrt{\lambda_\pm}$ determined by the %eigenvalues of $G_{ij}$:
%\bea
%\lambda_{\pm}=\frac{1}{2}\left(G_{11}+G_{22}\right)\pm \frac{1}{2}\sqrt{(G_{11}-%G_{22})^2+4G_{12}^2}.\eea
%which might be possible as the ratio $F_\xi/\sqrt{\lambda_\pm}$ involves tunable continuous parameters.
%In this regard, a particularly interesting scenario has been proposed recently in %\cite{Fonseca}, 

Our first example is a simple model involving two axions $\vec\theta=(\theta^1,\theta^2)$ 
and single $U(1)_A$ gauge boson, which was discussed recently in \cite{Fonseca:2019aux}. 
The model has a simple form of kinetic metric:
\bea
G=\left(\begin{array}{cc}  \ f_1^2\ & \ 0\ \\ 0\ & \ f_2^2\ \end{array}\right)
\eea
%G_{11}=f_1^2,\quad  G_{22}=f_2^2,\quad G_{12}=0,\eea
and the $U(1)_A$ gauge charge of $\vec\theta$:
% of the gauge field $A_\mu$, and the axions $\vec \theta(x)\equiv (\theta^1(x), \theta^2(x))$:
\bea
\label{model2}
%U(1)_A:\,\,\, A_\mu\to A_\mu  + \partial_\mu\Lambda,\quad \vec \theta =(\theta^1,\theta^2)  \,\to\, \vec\theta +\vec k\, \Lambda\,\,\,\,\, \mbox{with}\,\,\,\,\,
\vec k =(1,1).
\eea
For this model, we will examine 
 the reparameterization-invariant physical quantities such as the axion field range, the axion 1PI amplitude to gauge bosons, and the axion effective potential induced by non-perturbative gauge dynamics, which can be determined by the reparameterization-dependent  axion couplings in the model.

For the above $U(1)_A$ charge vector $\vec k$, the corresponding $\vec{\tilde k}$, $\vec\ell$ and $\vec{\tilde \ell}$ satisfying (\ref{vec_dual})  can be easily found to be\bea
\vec{\tilde k}=(1,-1), \quad \vec\ell =(1,0), \quad \vec{\tilde \ell}=(0,1).\eea 
We may take different $\vec\ell$ and $\vec{\tilde\ell}$ given by (\ref{degenerate}), but all physical consequences should be the same. 
According to our results in the previous section, 
the gauge-invariant axion combination $\xi$ and the
gauge-variant $\zeta$ eaten by the $U(1)_A$ gauge boson are
given by  
\bea
 \xi &=& M_\xi \sum_i \tilde k_i \theta^i =M_\xi (\theta^1-\theta^2), \nonumber\\
\zeta &=& M_A^{-2}\sum_{ij} G_{ij} k^i \theta^j=\frac{f_1^2\theta^1 +f_2^2 \theta^2}{f_1^2+f_2^2},\eea
where
\bea
M_A^2 &=& \sum_{ij} G_{ij}k^ik^j = f_1^2 + f_2^2, \nonumber \\
M_\xi^2 &=& \frac{1}{\sum_{ij} (G^{-1})^{ij}\tilde k_i\tilde k_j}=\frac{f_1^2 f_2^2}{f_1^2 + f_2^2},
\eea
and the $U(1)_A$ gauge boson mass and the field range of $\xi$ are determined as
$gM_A$ and $\Delta\xi=2\pi M_\xi$, respectively.
Equivalently, the original angular axions $\vec\theta$ can be decomposed as 
\bea
\vec \theta =  \vec k\, \zeta +  \Big( \vec \ell+ \Gamma(G,\vec k) \,\vec k\Big) \frac{\xi}{M_\xi},\eea
where
\bea
\Gamma(G,\vec k) = \frac{\sum_{ij}(G^{-1})^{ij}\tilde \ell_i\tilde k_j}{\sum_{ij}(G^{-1})^{ij}\tilde k_i\tilde k_j}=-\frac{f_1^2}{f_1^2 + f_2^2}.
\eea

%Note that the $U(1)_A$ gauge boson mass and the axion field range  are determined by the axion kinetic metric $G_{ij}$ and
%the $U(1)_A$ charge $k^i$ without any reference to the other parts of the model. 

Let us consider the possible axion couplings to gauge and matter fields in this model.
In \cite{Fonseca:2019aux}, it was noticed that this model can give a highly suppressed coupling of $\xi$ to non-Abelian
 gauge bosons
in the parameter limit\footnote{As was discussed in \cite{Fonseca:2019aux}, this scale hierarchy
 can be achieved by either a warped extra dimension or nearly conformal 4D dynamics in the underlying UV theory.}
\bea
f_1 \gg  f_2.\eea
This observation is based on the coupling
\bea
\label{axion_gauge}
\frac{1}{32\pi^2}\vec r\cdot\vec \theta\, G^a_{\mu\nu}\tilde G^{a\mu\nu}\quad \mbox{with}\quad
\vec r = (1,0),
\eea
which results in
\bea
\frac{1}{32\pi^2}\frac{\xi}{f_\xi} G^a_{\mu\nu}\tilde G^{a\mu\nu},\
\eea
where
\bea
\frac{1}{f_\xi} = \frac{1}{M_\xi} \Big( \vec r\cdot\vec \ell+  \Gamma(G,\vec k)\, \vec r\cdot \vec k  \Big)
%=\Big(1+\Gamma\Big)\frac{1}{M_\xi}
=\frac{f_2^2}{f_1^2+f_2^2}\frac{1}{M_\xi}.
%=\frac{f_2}{f_1\sqrt{f_1^2+f_2^2}}
\eea
Then in the limit $f_1\gg f_2$, the effective coupling $1/f_\xi$ 
is much smaller than the size ($\sim 1/M_\xi$) one would naively expect from the axion field range $\Delta\xi=2\pi M_\xi$.
On the other hand, the above expression of $1/f_\xi$ shows that the big suppression of $1/f_\xi$ relative to $1/M_\xi$ is possible only when   the coupling (\ref{axion_gauge}) is {\it not} gauge-invariant by itself, i.e. only when
$\vec r\cdot\vec k\neq 0$. Note that if $\vec r\cdot\vec k=0$, then $1/f_\xi = (\vec r\cdot\vec \ell) /M_\xi$  is an integer multiple of $1/M_\xi$ as expected.
If $\vec r\cdot\vec k\neq 0$, the model should include  gauge-charged chiral fermions whose mixed $U(1)_A\times [SU(N_c)]^2$ anomaly cancels the $U(1)_A$ variation of the coupling (\ref{axion_gauge}). In the presence of such chiral fermions, the coupling $1/f_\xi$ varies under the $\xi$-dependent phase rotation of fermion fields, implying that 
the suppression of $1/f_\xi$ is an artifact 
of the particular choice of  field basis, so needs more careful interpretation\footnote{In fact, the discussion of \cite{Fonseca:2019aux} relies on the axion potential derived in
\cite{Shiu:2018unx}, which is not compatible with our result in the previous section. This discrepancy arises from that 
the fermion bilinear condensation $\langle \psi\chi\rangle$ is treated  as a field-independent constant in the discussion of axion potential in \cite{Shiu:2018unx}.  If one takes into account the correct field-dependence of
the fermion bilinear condensation, i.e. $\langle \psi\chi\rangle\propto e^{i\eta/f_\eta}$ for the composite meson field $\eta$, the resulting axion potential becomes compatible with our results.
}.

To see this, let us introduce the required fermions which can cancel the $U(1)_A$ variation of (\ref{axion_gauge}): 
%i.e. $\psi_I$ ($I=1,2$) with the $SU(N_c)\times U(1)_A$ gauge charges
\bea
\psi= ( N_c, q_{\psi}), \quad \chi =( \bar{N}_c, q_{\chi}), 
%\quad \phi=(  N_c, q_\phi) 
\eea
%to cancel the $U(1)_A$ variation of the coupling  (\ref{axion_gauge}),
% as well as a complex scalar $\phi$,  
where $N_c$ and $\bar N_c$ denote the fundamental and anti-fundamental representation of $SU(N_c)$,  and $q_{\psi,\chi}$ are the $U(1)_A$ charges of $\psi,\chi$. Then the generic axion couplings to gauge and matter fields take the form 
\bea
\label{axion_int}
{\cal L}_{\rm int}
&=&\frac{1}{32\pi^2}\vec r\cdot\vec \theta\,  G^a_{\mu\nu}\tilde G^{a\mu\nu} -
\left(\mu e^{i\vec n\cdot\vec\theta}\psi\chi  + {\rm  h.c.}\right)
%+\,\vec \kappa_\phi\cdot\partial_\mu \vec \theta \,(\phi^*D_\mu\phi- \phi D_\mu\phi^*)
+ \sum_{\psi_I=\psi,\chi}\vec \kappa_{\psi_I}\cdot D_\mu \vec \theta\,\bar\psi_I\sigma^\mu \psi_I\nonumber \\
&=& \frac{c_g}{32\pi^2}\frac{\xi}{M_\xi}  G^a_{\mu\nu}\tilde G^{a\mu\nu} - \left(\mu e^{ic_\mu \xi/M_\xi}\psi\chi+{\rm h.c}\right)+\frac{\partial_\mu\xi}{M_\xi}\sum_{\psi_I=\psi,\chi} c_{\psi_I}\bar\psi_I\sigma^\mu \psi_I,
\eea 
 where the gauge-variant $\zeta$ is integrated out in the latter expression. 
The $U(1)_A$ invariance of the model requires
\bea
q_{\psi}+q_{\chi}= \vec r\cdot\vec k= 1, \qquad
\vec n\cdot \vec k =n_1+n_2=q_{\psi}+q_{\chi} = 1,
\eea
where the first condition is for the cancellation of the $U(1)_A\times [SU(N_c)]^2$ anomaly.
%For the specific model with under consideration, this leads to
%\bea
%n_1+n_2=q_{\psi_1}+q_{\psi_2}=1,\eea
Then the four axion couplings $c_g, c_\mu$ and $c_{\psi_I}$ ($\psi_I=\psi,\chi$) are given by
\bea
c_g &=&  \vec r\cdot\vec\ell+  \Gamma(G,\vec k)\, \vec r\cdot \vec k= \frac{f_2^2}{f_1^2+f_2^2},\nonumber \\
c_\mu &=&\vec n\cdot \vec\ell+  \Gamma(G,\vec k) \, \vec n\cdot \vec k =\frac{n_1 f_2^2- (1-n_1) f_1^2}{f_1^2+f_2^2}, \nonumber \\
c_{\psi_I}&=&\vec\kappa_{\psi_I}\cdot \vec\ell+   \Gamma(G,\vec k) \, \vec\kappa_{\psi_I}\cdot \vec k  
= \frac{\kappa_{\psi_I 1} f_2^2-\kappa_{\psi_I 2} f_1^2}{f_1^2+f_2^2}.
\eea
We can now consider the two parameter family of $\xi$-dependent field redefinition:
\bea
\label{repara}
\psi \,\to\, e^{ix_{\psi}\xi/M_\xi} \psi, \quad \chi \,\to\, e^{ix_{\chi}\xi/M_\xi} \chi,
%\quad \phi\,\to\, e^{i x_\phi \xi/M_\xi}\phi
\eea
under which the axion couplings vary as 
\bea
\label{coupling_change}
c_g \,\to\,  c_g +x_{\psi} +x_{\chi}, \quad
c_\mu \,\to\, c_\mu +x_{\psi}+x_{\chi},
\quad c_{\psi_I}\,\to\, c_{\psi_I} -x_{\psi_I}\,\, \,(\psi_I=\psi,\chi).
\eea
Here the change of $c_g$ is due to the anomalous variation of the path integral measure of $\psi_I$, while the change of
the derivative couplings $c_{\psi_I}$  originates from the kinetic terms of $\psi_I$. 
In the previous section, we considered a particular field redefinition with
\bea
x_{\psi_I} = -q_{\psi_I} \Gamma(G,\vec k) \quad (\psi_I=\psi,\chi),\eea
%=\frac{q_{\psi_I}f_1^2}{f_1^2+f_2^2},\eea
after which all non-derivative couplings of $\xi$ are quantized to be manifestly consistent with the axion periodicity $\xi\equiv \xi +2\pi M_\xi$. Indeed, we find that the corresponding $c_g$ and $c_\mu$ have integer values as
\bea \label{c_g_mu}
c_g = \vec r\cdot\vec\ell=1,\quad
c_\mu =\vec n\cdot\vec \ell= n_1,\eea
while the derivative couplings are shifted as
\bea
c_{\psi_I}= 
\vec \kappa_{\psi_I} \cdot \vec\ell  + 
    \Gamma(G, \vec k)(\vec\kappa_{\psi_I} \cdot \vec k + q_{\psi_I} ) .
\eea

At any rate, all physical consequences of the axion coupling
(\ref{axion_int}) should be invariant under the field redefinition  (\ref{repara}), and therefore
determined by the following two reparameterization-invariant coupling combinations:
\bea
\label{ri_combination}
  c_g-c_\mu &=& (\vec r -\vec n)\cdot  \vec \ell\, = \, 1-n_1, \nonumber \\
  c_{\psi}+c_{\chi}+c_\mu &=& (\vec \kappa_{\psi}+ \vec \kappa_{\chi} +\vec n)\cdot \Big( \vec\ell+  \Gamma(G,\vec k)\vec k \Big) \nonumber\\
 && \hskip -2.5cm =\,\frac{ f_2^2}{f_1^2+ f_2^2}\Big( (\vec\kappa_{\psi})_1 + (\vec \kappa_{\chi})_1 + n_1\Big) 
  -\frac{f_1^2}{f_1^2+ f_2^2} \Big((\vec \kappa_{\psi})_2 + (\vec \kappa_{\chi})_2 + (1-n_1)\Big).  \eea
%Note that $c_g- c_\mu$ is also invariant under the change of $\ell$ as (\ref{degenerate}) 
%because  $\vec r - \vec n = (1-n_1 ,-n_2) = n_2(1, -1)= n_2\tilde k$ as discussed in (\ref{consistency}).  
For the specific model under discussion,  $c_g=M_\xi/f_\xi \simeq f_2^2/f_1^2 \ll 1$ in the limit $f_2\ll f_1$. However the associated reparameterization-invariant combination $c_g - c_\mu$, which is relevant for the generation of the axion effective potential, 
has an integer value which is independent of $f_{1,2}$.  In regard to this, one may make an analogy with the QCD.  The
 combination $c_g - c_\mu$ is an analogue of
the reparameterization-invariant QCD angle $\bar\theta =\theta_{\rm QCD}+ \arg\det (M_q)$, where $M_q$ is the light quark mass matrix, while the basis-dependent $c_g$ (or $1/f_\xi = c_g/M_\xi$) corresponds to the bare vacuum angle $\theta_{\rm QCD}$ whose physical consequences always appear through the invariant combination $\bar\theta$.
 In the following, we evaluate the axion effective potential and the axion 1PI amplitude to gauge bosons to confirm that they are indeed determined by the above two reparameterization-invariant parameter combinations.

\begin{figure}[t] %htbp]
\begin{center}
\includegraphics[width=11cm]{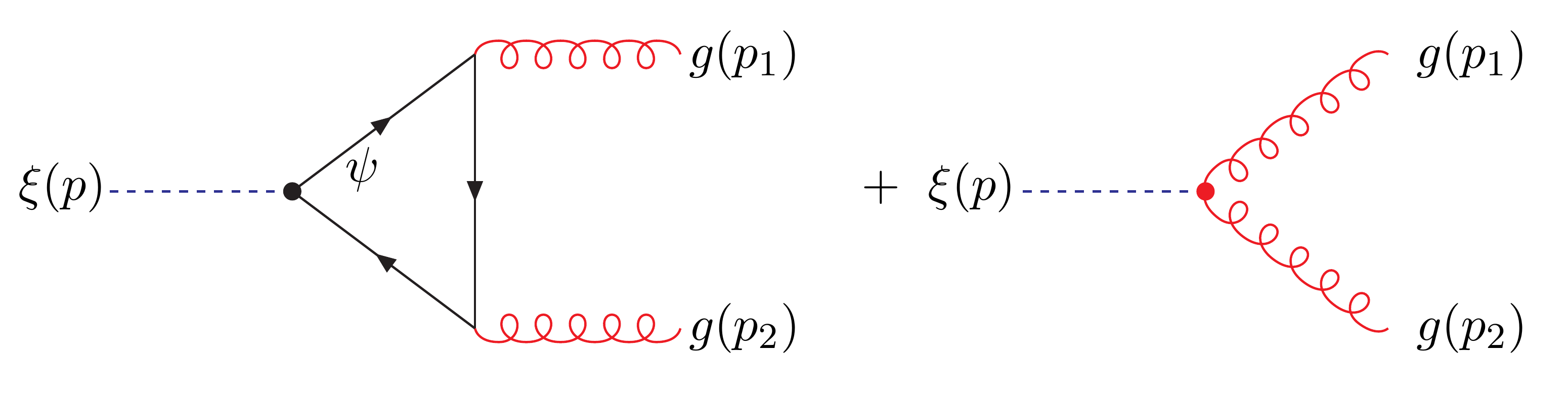}
\caption{Feynman diagrams for the 3-point 1PI amplitude of axion ($\xi$) and $SU(N_c)$ gauge bosons ($g$). The fermion loop involves either the derivative couplings $c_{\psi_i}$ or the non-derivative coupling $c_\mu$.}
\label{fig:feynman_1PI}
\end{center}
\end{figure}

Let us first consider the 3-point 1PI diagram of axion and $SU(N_c)$ gauge bosons   (Fig.~\ref{fig:feynman_1PI}) for the external momenta $|p_i|\gg   \Lambda_{SU(N_c)}$,
where $\Lambda_{SU(N_c)}$ denotes the confinement scale of the $SU(N_c)$ gauge interactions. It is straightforward to find that at one-loop approximation the amplitude   is given by 
 \bea
 \hskip -0.6cm
{\cal A}_{\xi GG} = \frac{ i\alpha_s}{2\pi M_\xi } \epsilon^{\mu\nu\rho\sigma}\epsilon_{1\mu}\epsilon_{2\nu}p_{1\rho}  p_{2\sigma} \Big[(c_g-c_\mu) + (c_{\psi}+c_{\chi}+c_\mu)
F(p_1,p_2;\mu)\Big],\eea
where $p_{i\mu}$ and $\epsilon_i^\mu$ are the 4-momenta and polarization vectors of the two external gauge bosons.
The loop function $F$ is given by 
\bea
F(p_1,p_2;\mu) = 1 - \int_0^1 dx \int_0^{1-x} dy \frac{2 \mu^2}{\mu^2 - ( p_1^2\, x(1-x) + p_2^2\, y(1-y) + 2 p_1\cdot p_2\, xy)},
\eea
which has the limiting behaviour
%When the gluons are on mass-shell, i.e. $p_1^2 = p_2^2=0$,  $F$ can be expressed analytically as the function of $\tau \equiv %4 \mu^2 / p^2$ as
\bea
\label{F(x)}
F(p_1, p_2: \mu) = \left\{\begin{array}{ll}  
-\frac{p_1^2+p_2^2+p^2}{12\mu^2}+{\cal O}\left(\frac{p^4}{\mu^4}\right)  &\quad {\rm for}\ \,\,  p_1^2\sim p_2^2\sim p^2 \ll \mu^2 \\
1+ {\cal O}\left(\frac{\mu^2}{p^2}\right) &\quad {\rm for}\ \,\, p_1^2\sim p_2^2\sim p^2 \gg \mu^2,
\end{array} \right. 
\eea
where $p^\mu=-(p_1^\mu+p_2^\mu)$ is the axion 4-momentum.
The above result shows that the 1PI amplitude ${\cal A}_{\xi GG}$ is determined indeed by the two  reparameterization-invariant combinations of (\ref{ri_combination}).
One can see also that in high energy limit with $p^2\gg \mu^2$, we have  $F \simeq 1$ and therefore  
  ${\cal A}_{\xi GG}$ is determined mostly by the reparameterization-invariant parameter combination $c_g + c_{\psi} + c_{\chi}$. On the other hand, in the heavy fermion limit with $\mu^2\gg p^2$,  $F={\cal O}(p^2/\mu^2)\ll 1$ and then  ${\cal A}_{\xi GG}$ is determined mostly by the other reparameterization-invariant combination $c_g- c_\mu$.
  
Although we literally call  $c_g$ and $c_\mu$ non-derivative couplings, 
 they can be regarded as derivative couplings in perturbation theory
 since  $G^a_{\mu\nu}\tilde G^{a\mu\nu}$ is a total divergence and $c_\mu$ can be rotated away into $c_g$ and  $c_{\psi,\chi}$ by an appropriate  $\xi$-dependent field redefinition (\ref{repara}). As a result,  all perturbative amplitudes induced by the axion couplings $c_g, c_\mu$ and $c_{\psi,\chi}$ are vanishing in the limit when the external axion momentum $p$ becomes zero. This can be easily understood by the continuous PQ symmetry  
\bea
U(1)_{PQ}: \quad \xi\rightarrow \xi+ c\,M_\xi, \quad \psi\chi \rightarrow e^{-i cc_\mu}\psi\chi \quad (c=\mbox{constant}),\eea
 which is an exact symmetry in perturbation theory in our case.
 
Of course, the above PQ symmetry can be explicitly broken by  the $U(1)_{PQ}\times [SU(N_c)]^2$ anomaly through
non-perturbative effects such as the
$SU(N_c)$ instantons with \bea
\int d^4x \, G^a_{\mu\nu}\tilde G^{a\mu\nu} \neq 0,\eea 
and also possibly by non-perturbative quantum gravity effects.
Including such nonperturbative $SU(N_c)$ dynamics, $c_g$ and $c_\mu$, more precisely the reparameterization-invariant combination $c_g-c_\mu$, can be regarded as  genuine non-derivative coupling which can generate a non-trivial axion potential. 
%As a physical quantity which iscan consider  the axion potential generated by non-perturbative $SU(N_c)$ the non-derivative %interactions in (\ref{axion_int}). 
To have a small parameter which allows a systematic expansion of the generated axion potential, let us assume that
$\mu/\Lambda_{SU(N_c)}\ll 1$. Then, at scales around $\Lambda_{SU(N_c)}$, the light fermions $\psi,\chi$ form a bilinear condensation which can be parametrized as 
\bea
\langle \psi\chi\rangle = \Lambda_{\psi}^3  e^{i\eta/f_\eta},\eea
where $\eta$ is a composite meson and $\Lambda_\psi\sim f_\eta \sim \Lambda_{SU(N_c)}$. The behavior (\ref{coupling_change}) of axion couplings under the fermion field redefinition (\ref{repara}) implies that the meson potential should be
 invariant under the following spurion transformation of the field and parameters:
\bea
&& \hskip 1cm \frac{\eta}{f_\eta} \,\to\, \frac{\eta}{f_\eta} +(x_{\psi}+x_{\chi})\frac{\xi}{M_\xi},\nonumber \\ 
&& c_g \,\to\,  c_g  + x_{\psi} + x_{\chi}, \quad
c_\mu \,\to\, c_\mu +x_{\psi}+x_{\chi}.
\eea
This suggests that 
%the meson potential   in powers of $\mu/\Lambda_{SU(N_c)}\ll 1$, one then finds 
 \bea 
 \label{meson_potential}
V(\xi, \eta) = U\Big(\frac{\eta}{f_\eta}+ c_g\frac{\xi}{M_\xi}\Big) - \mu \Lambda_{\psi}^3 \cos\Big(\frac{\eta}{f_\eta} - c_\mu\frac{\xi}{M_\xi}\Big) +{\cal O}(\mu^2\Lambda_{SU(N_c)}^2),\eea
where $U(\theta)={\cal O}(\Lambda_{SU(N_c)}^4)$ is a periodic function of $\theta$ which has a global minimum at $\theta=0$. Note that the precise form of $U(\theta)$ depends on the details of non-perturbative $SU(N_c)$ dynamics, while 
the next term is unambiguously determined to be a simple cosine function.
Without knowing the detailed form of $U(\theta)$, we can integrate out $\eta$ by minimizing $U(\theta)$ ($\theta\approx 0$), which results in 
% to get the effective potential of the light axion $\xi$ at leading order in $\mu/\Lambda_{SU(N_c)}$:
\bea
\frac{\eta}{f_\eta} = -c_g\frac{\xi}{M_\xi} +{\cal O}\Big(\frac{\mu}{\Lambda_{SU(N_c)}}\Big).\eea
Inserting this solution to the potential (\ref{meson_potential}), one finds the axion effective potential is given by
\bea
V_{\rm eff}(\xi) =-\mu \Lambda_{SU(N_c)}^3\cos\Big((c_g -c_\mu)\frac{\xi}{M_\xi}\Big) +{\cal O}(\mu^2\Lambda_{SU(N_c)}^2).\eea
Since the basis-independent  $c_g - c_\mu$  has an integer value,  this axion potential is manifestly consistent with the axion periodicity $\xi\equiv \xi+2\pi M_\xi$ regardless of the value of $c_g$ which can be highly suppressed in some particular field basis. 
%We also point out that in the field basis to yield (\ref{c_g_mu}), it is obvious 
%that $c_g- c_\mu$ is an integer because each coefficient is the integer. 

\subsection{Models of multiple axions with clockwork-type $U(1)$ gauge charges}

Our next example is  a model of  $N$ axions $\vec\theta=(\theta^1,\theta^2, \cdots, \theta^N)$ which have the St\"uckelberg mixing with
 $(N-1)$ $U(1)$ gauge bosons $A_\mu^\alpha$ ($\alpha=1,2,..., N-1$).
 The  $U(1)$ gauge charges of $\vec\theta$ take the clockwork form \cite{Bonnefoy:2018ibr}: 
\bea 
\label{cw_charge}
\vec k_1 = (1, -q, 0, 0, \cdots,0), \  \vec k_2 = (0, 1, -q, \cdots, 0),\  \cdots, \  \vec k_{N-1}= (0,0, \cdots, 1, -q)
\eea
%\bea
%k^i_\alpha =\delta^i_\alpha  - q \delta^{i-1}_{\alpha}.
%\eea  
for an integer $q>1$. Then the corresponding $\vec{\tilde k}$, $\vec\ell$ and $\vec{\tilde \ell}^\alpha$ are found to be 
%\bea 
%\tilde k_i = q^{N-i},\quad \ell^i = \delta^i_N,
%\quad \tilde\ell_i^\alpha = \left\{ 
%\begin{array}{ll} 
%q^{\alpha-i}  & \textrm{for $\alpha\geq i $} \\
% 0 & \textrm{for $\alpha<i $}\end{array}\right.,
%\eea 
%which result in
\bea
%\vec k_{N-1} = (0, 0,\cdots 1, -q), \quad 
&& \hskip 2cm  
\vec{\tilde k}=(q^{N-1}, q^{N-2},\cdots, q, 1),\quad \vec \ell = (0,0,\cdots ,0, 1),
 \nonumber\\
&& \vec{\tilde \ell}^1 = (1,0,\cdots,0, 0), \quad \vec{\tilde \ell}^2 = (q,1,\cdots,0, 0), \quad 
\cdots,\quad
\vec{\tilde \ell}^{N-1} = (q^{N-2}, q^{N-3}, \cdots, 1,0).\nonumber 
\eea 
For simplicity, we assume all $U(1)_\alpha$ gauge couplings are universal, $g_\alpha=g$, and take the most simple form of the axion kinetic metric: \bea
G_{ij} = \delta_{ij} f^2.\eea
Then the gauge boson mass matrix is given by
\bea \label{mass_vector}
g^2M_A^2 = g^2\sum_{ij} G_{ij} k^i k^j = 
g^2 f^2 \left(\begin{array}{ccccc}  
 1+q^2  & - q & 0 &\cdots & 0 \\
 -q & 1+q^2 & -q &\cdots & 0 \\ 
 \vdots   & \vdots & \ddots & \vdots  &-q \\ 
 0 & 0& \cdots  &-q & 1+q^2 
 \end{array}\right), 
\eea  which results in the mass eigenvalues 
\bea 
\left(M_A^{(n)}\right)^2 = g^2 f^2 \left(1+ q^2 - 2q \cos\frac{n \pi }{N}\right) 
\quad (n=1,2,\cdots, N-1).
\eea 
As anticipated in the previous section, the axion decay constant of $\xi$, which is defined by the periodicity $\xi\equiv \xi+2\pi M_\xi$, is exponentially reduced as 
\bea  
M_\xi= \frac{1}{\sqrt{\sum_{ij} (G^{-1})^{ij} \tilde k_i \tilde k_j} } = 
\frac{f}{{||\tilde{k}||}} = \sqrt{\frac{q^2- 1}{q^{2N} - 1}}f\,\sim\, \frac{f}{q^{N-1}},
\eea 
which is consistent with the behavior (\ref{m_xi_behavior}) in the limit $N\gg 1$ as the root mean square  of the
 $U(1)$ charges   is estimated as  $(k^i_{r\alpha})_{\rm rms}\sim q/\sqrt{N}$.

 However, having the field range $M_\xi \ll f$ does not mean
that the couplings of $\xi$ are of the order of $1/M_\xi \gg 1/f$. The gauge-invariant axion
$\xi$ interacts with other fields through the wavefunction mixing between $\xi$ and the original angular axions $\theta^i$. 
Then from the decomposition 
\bea \label{theta_CW}
\theta^i =\sum_{\alpha=1}^{N-1} k^i_\alpha \zeta^\alpha +   
\frac{q^2 -1}{(q^N - q^{-N})q^{i}}\frac{\xi}{M_\xi}  
= \sum_{\alpha=1}^{N-1} k^i_\alpha \zeta^\alpha +   \Big( \ell^i + \sum_{\alpha=1}^{N-1}\Gamma^\alpha\, 
k^i_\alpha \Big) \frac{\xi}{M_\xi},
\eea 
where 
\bea \label{gamma_CW}
\Gamma^\alpha= \frac{q^\alpha - q^{-\alpha}}{q^{N}-q^{-N}},
\eea
one immediately finds that the wavefunction mixing is given by
\bea
\langle \theta^i|\xi\rangle \,\sim\, \frac{1}{q^{i-1}}\,\frac{1}{f}.\eea
As a consequence, unless one makes a $\xi$-dependent field redefinition which may change the characteristics of axion couplings,
all couplings of $\xi$ are of the order of $1/f$ or smaller if the couplings of $\theta^i$ are of order unity or smaller.

% is much suppressed than 
%$1/M_\xi$ for all $\theta^i$. 
%This implies that generating a scalar potential of $\xi$ would be nontrivial as discussed in Sec.~\ref{Large_N}.
%Because our framework is effective theory, 
%applying naive dimensional analysis, the UV cut-off can be taken  as $\Lambda_{UV} = 4\pi f$.  
%Then 
%$U(1)$ symmetries allow us to write down a bare scalar potential of $\xi$ 
%through the interaction term like 
%\bea
%\Delta {\cal L}_{\rm bare} = -\epsilon_{\rm bare} \Lambda_{UV}^4 e^{i \tilde k_i\theta^i } + {\rm h.c.} \to    V_{UV}(\xi)= %2\epsilon_{\rm bare}\Lambda_{UV}^4 \cos\Big(\frac{\xi}{M_\xi}\Big). 
%\eea 
%Here $\epsilon_{\rm bare}$ is not predictable, but 
%its size might be quite small,  
%because the coupling $\tilde k_i = q^{N-i}$ is unusually large for $N\gg 1$, $q>1$. 
%In this paper, instead of focusing on UV physics to generate bare potentials of $\xi$, 
%we will focus on the calculable part of scalar potentials whose contributions are dominated by IR physics with natural size of %the couplings. 

%The right side expression of the second equality in (\ref{theta_CW}) and the field basis of (\ref{field_redef}) are useful when we explicitly calculate the scalar potential at scales lower than $M_\xi$, because the derivative interactions are irrelevant at low scales, then the $2\pi$ periodicity of $\xi/M_\xi$ is manifest. In the following subsections, we discuss some explicit examples. 

%\subsubsection{An example with charged fermions}
To examine the axion couplings more explicitly, let us introduce $SU(N_c)$ gauge fields $G^a_\mu$ and $N_f$ pairs of chiral fermions ($\psi_I,\chi_I$) ($I=1,2,..,N_f)$ whose 
$SU(N_c)\times \prod_\alpha U(1)_\alpha$ gauge charges are given by
\bea
\psi_I = (N_c, q_{\psi_I \alpha}),\quad 
\chi_I = (\bar N_c, q_{\chi_I\alpha}). 
\eea 
Those gauge and matter fields can couple to $\theta^i$ as 
\bea\label{int_CW}
&& \frac{1}{32\pi^2}\Big(\sum_i r_i\theta^i\Big) G^a_{\mu\nu}\tilde G^{a\mu\nu}
 - \Big(\sum_{I}\mu_{I}e^{i\sum_i n^{I}_ i\theta^i}\psi_I\chi_I+{\rm h.c.}\Big) \nonumber\\
 &&\hskip 1cm +\,\sum_{I,i}
D_\mu\theta^i \Big(\kappa^{I}_{\psi i } \bar\psi_I\sigma^\mu\psi_I
+ \kappa^{I}_{\chi i}\bar\chi_I\sigma^\mu\chi_I\Big),\eea
where the invariance under $\prod_\alpha U(1)_\alpha$ requires
\bea \label{gauge_inv}
\sum_i r_i k^i_\alpha = \sum_{I} (q_{\psi_I \alpha} + q_{\chi_I \alpha}),\quad 
\sum_i n^{I}_ik^i_\alpha = q_{\psi_I \alpha} + q_{\chi_I\alpha}. 
\eea  
Here for simplicity we consider only the flavour-diagonal axion couplings. 
After integrating out the massive $U(1)_\alpha$ gauge fields $A^\alpha_\mu-\partial_\mu\zeta^\alpha$, one finds  
the low energy couplings of $\xi$ given by 
\bea 
&& \frac{c_g}{32\pi^2}\frac{\xi}{M_\xi}G^{a\mu\nu}\tilde G^a_{\mu\nu} - \Big(\sum_{I}\mu_{I} e^{ic^{I}_\mu \xi/M_\xi}\psi_I\chi_I+{\rm h.c.}\Big) \nonumber\\
&&\hskip 1cm +\,\frac{\partial_\mu\xi}{M_\xi}\sum_{I}  
\left(c_{\psi}^{I}\bar\psi_I\sigma^\mu \psi_I+c_{\chi}^{I}\bar\chi_I\sigma^\mu \chi_I \right), 
\eea   where 
%the couplings $c_g$, $c_\mu^{IJ}$, $c_{\psi_I}$, $c_{\chi_I}$ are 
\bea \label{coeff_1}
c_g =   \frac{q^2 - 1}{q^N- q^{-N}} \sum_i \frac{r_i}{q^{i}},\qquad 
%= \vec r\cdot\vec\ell+  \Gamma^\alpha\, \vec r\cdot \vec k_\alpha,\nonumber \\
c_\mu^{I} =  \frac{q^2 - 1}{q^N- q^{-N}} \sum_i \frac{n_i^{I}}{q^{i}},
%= \vec n^{IJ}\cdot\vec\ell+  \Gamma^\alpha\, \vec n^{IJ}\cdot \vec k_\alpha, 
\nonumber \\
c_{\psi}^{I}=
  \frac{q^2 - 1}{q^N- q^{-N}} \sum_i \frac{\kappa_{\psi i}^{I}}{q^{i}} , \qquad
  %= \vec\kappa_{\psi }^{IJ}\cdot \vec\ell+   \Gamma^\alpha \, \vec\kappa_{\psi }^{IJ}\cdot \vec k_\alpha, \nonumber\\
c_{\chi}^{I} = 
  \frac{q^2 - 1}{q^N- q^{-N}} \sum_i \frac{\kappa_{\chi i}^{I}}{q^{i}}. 
%= \vec\kappa_{\chi }^{IJ}\cdot \vec\ell+   \Gamma^\alpha \, \vec\kappa_{\chi }^{IJ}\cdot \vec k_\alpha,  
\eea 
Hence in this prescription, 
%due to  $\langle \theta^i|\xi\rangle \lesssim  {\cal O}(1/q^N M_\xi)$, 
the effective axion couplings
   $c_g, c_\mu^{I}, c_\psi^{I}$ and  $c_\chi^{I}$ are manifestly suppressed by  $1/q^{N}\ll 1$ relative to the original axion couplings  $ r_i, n_i^{I}, \kappa_{\psi i}^{I}, \kappa_{\chi i}^{I}$, and therefore  
all couplings of $\xi$  are of the order of $1/f\sim 1/q^{N-1} M_\xi$ or smaller as long as  the original couplings of $\theta^i$ are of order unity or smaller.
On the other hand, in the new field basis after the field redefinition  (\ref{field_redef}), one finds 
% couplings  are given by
\bea \label{coeff_2}
&& \hskip 2cm 
 c_g = \sum_i  r_i \ell^i =r_N, \quad  c_\mu^{I} = 
\sum_i n^{I}_i\ell^i = n_N^{I},  \nonumber\\
&& c_{\psi}^{I} = 
\kappa_{\psi N}^{I} + \sum_\alpha\Gamma^\alpha (q_{\psi_I \alpha}+\sum_i\kappa_{\psi i }^{I}k^i_\alpha  ), \quad c_{\chi}^{I} = \kappa_{\chi N}^{I} + \Gamma^\alpha (q_{\chi_I \alpha} +\sum_i \kappa_{\chi i }^{IJ}k^i_\alpha ),
\eea    
so the couplings do not reveal a suppression by $1/q^N$.
%  relative to the original axion couplings  $ r_i, n_i^{I}, \kappa_{\psi i}^{I}, \kappa_{\chi i}^{I}$. 
This means that the characteristic size of axion couplings is so different in the two different field bases. 
On the other hand, all physical consequences of the model should be independent of the choice of field basis.
This indicates that one needs to be careful when examine the physical consequences of the axion couplings
in the new field basis after the field redefinition  (\ref{field_redef})
as there can be a fine cancellation among the contributions from different couplings.

To avoid a confusion due to the basis-dependent feature of the couplings, let us consider the basis-independent (reparameterization-invariant) combinations of axion couplings. Although we have introduced $3N_f+1$ axion couplings, all of their physical consequences can be described by the $N_f+1$ combinations of couplings:
\bea
&& \tilde c_g\equiv c_g -\sum_I c_\mu^I= \sum_i \Big(r_i-\sum_I n^I_i\Big)\ell^i =r_N -\sum_I n_N^I, \nonumber \\
&& \tilde c_I\equiv c^I_\psi+c^I_\chi+c_\mu^I =\frac{q^2 - 1}{q^N- q^{-N}} \sum_i\frac{\kappa_{\psi i}^I+\kappa_{\chi i}^I +n_i^I}{q^i},
 \eea
which are
invariant under the field redefinition
\bea
\psi_I\rightarrow e^{ix_{\psi_I}\xi/M_\xi}\psi_I, \quad  \chi_I\rightarrow e^{ix_{\chi_I}\xi/M_\xi}\chi_I.\eea
For instance, the perturbative 3-point 1PI amplitude of $\xi$ to $SU(N_c)$ gauge fields is given by  
 \bea
 \hskip -0.6cm
{\cal A}_{\xi GG}  = \frac{ i\alpha_s}{2\pi M_\xi } \epsilon^{\mu\nu\rho\sigma}\epsilon_{1\mu}\epsilon_{2\nu}
p_{1\rho}  p_{2\sigma} \Big[ \tilde c_g + \sum_I \tilde c_I
F(p_1,p_2;\mu_I)\Big],\eea
where  $p_{i\mu}$ and $\epsilon_i^\mu$ are the 4-momenta and polarization vectors of the two external gauge bosons and the loop function $F$ is given by (\ref{F(x)}), 
while the axion potential generated by non-perturbative $SU(N_c)$ dynamics is determined by the integer-valued combination 
$\tilde c_g$  as
\bea
V_{\rm eff}\Big( \tilde c_g \frac{\xi}{M_\xi}\Big),\eea
where
$V_{\rm eff}(x)$ is a $2\pi$ periodic function of $x$, whose form is determined by more details of the model.   Note that
$\tilde c_I$ are continuous real numbers, while $\tilde c_g$ is integer-valued.

One might  be puzzled about that 
the basis-independent combinations
$\tilde c_I$ are suppressed by $1/q^N$ relative to the original axion couplings such as $\kappa_{\psi i}^I,\kappa_{\chi i}^I, n_i^I$, while
there is no such suppression for $\tilde c_g$. This suggests that the model should involve a structure yielding an exponentially large number of ${\cal O}(q^N)$ to have $\tilde c_g\neq 0$. Indeed 
the $U(1)_\alpha$ ($\alpha=1,2,..,N-1)$ gauge symmetries require such a structure and make it highly non-trivial to achieve $\tilde c_g\neq 0$.
To see this, 
let us note that the  gauge invariance condition (\ref{gauge_inv}) implies
\bea
\sum_i \Big(r_i-\sum_I n_i^I\Big)k^i_\alpha =0 \quad \mbox{for all}\,\,\,\alpha=1,2,...,N-1,\eea
and therefore $\vec r-\sum_I \vec n^I$ should be proportional to $\vec{\tilde k}$.
Combined with $\tilde c_g=r_N-\sum_I n_N^I$ from (\ref{coeff_2}), this determines $\vec r-\sum_I \vec n^I$ as
\bea
\label{condition_u1}
\vec r-\sum_I \vec n^I= \tilde c_g\vec {\tilde k} =\tilde c_g\,(q^{N-1}, q^{N-2},\cdots, q, 1).\eea
%In order to generate a non-trivial potential of $\xi$ from non-perturbative $SU(N_c)$ dynamics, one needs $\tilde c_g\neq 0$. %The above condition from the invariance under $\prod_\alpha U(1)_\alpha$ shows that 
Therefore $\tilde c_g$ can be non-zero 
in the limit $N\gg 1$
 {\it only when} the model parameters $r_i$ and/or  $\sum_I n^I_i$ are
 exponentially large as ${\cal O}(q^{N-i})$.     In other word, 
$\xi$ can get a non-trivial potential from  non-perturbative $SU(N_c)$ dynamics only in the extreme case involving exponentially large parameter and/or exponentially many degrees of freedom \cite{Bonnefoy:2018ibr}.
% which usually makes the model to have bad UV behaviour at scales above the St\"uckelberg scale \cite{dudas}. 
This reflects  that in the limit $N\gg 1$ the axion $\xi$ is so well protected by the $U(1)_\alpha$ ($\alpha=1,2,.., N-1$) gauge symmetries from getting a mass, so can be ultra-light in a natural manner.

%Let us close the discussion of the axion potential induced by $SU(N_c)$ dynamics  by providing a specific extreme model
%integer-valued model parameters $\vec r$ and $\vec n^I$ ($I=1,2,.., N_f$), 
%yielding $\tilde c_g\neq 0$, while satisfying the gauge invariance condition 
%(\ref{condition_u1}).
%Our example involves exponentially many gauge-charged fermion species $(\psi_I,\chi_I)$ with
% $N_f =  (q^N-1)/(q-1)$, and the model parameters 
% with $\vec r=0$ and
%\bea 
%\vec n^{I} &=& (1, 0,\cdots, 0)\quad {\rm for}\quad  I=1,\cdots, q^{N-1},\nonumber\\
%                 &=& (0, 1,\cdots, 0)\quad {\rm for}\quad I=1+ q^{N-1},\cdots, q^{N-2}+ q^{N-1},\cdots, \nonumber\\
%                 &=&(0,0,\cdots, 1)\quad  {\rm for}\quad  I = N_f= (q^N-1)/(q-1), 
%\eea 
%for which 
%$\tilde c_g= -1$.  
%For simplicity, we assume $\mu_{N_f} \ll \Lambda_{SU(N_c)}\ll \mu_I$ ($I=1,2,.., N_f-1$), where
%$\Lambda_{SU(N_c)}$ is the $SU(N_c)$ confinement scale. As the axion potential is generated mostly by the
%$SU(N_c)$ dynamics around $\Lambda_{SU(N_c)}$, one can integrate out the heavy fermion species $(\psi_I, \chi_I)$ ($I=1,2,.., %N_f-1$) and include also the effects of the condensation of the lightest fermion flavour $(\psi_{N_f}, \chi_{N_f})$:
% \bea 
% \langle \psi_{N_f}\chi_{N_f}\rangle = \Lambda^3 e^{ i \eta_{N_f}/f_\eta}.
% \eea
% Repeating the same analysis as in the previous subsection,
% we find the axion effective potential
% \bea 
% V_{\rm eff}(\xi) \simeq - \mu_{N_f N_f} \Lambda^3 \cos\Big(\frac{\xi}{M_\xi}\Big) +{\cal O}(\mu_{N_f N_f}^2 \Lambda^2).
% \eea 

As the above discussion suggests, one needs 
a  non-trivial engineering to generate an effective potential of $\xi$. We close this subsection by providing one example of such engineering.
Our example involves $N$ complex scalar fields $\phi_P$ 
with the $U(1)_\alpha$ gauge charges
\bea
q_{\phi_P \alpha} = \delta_{P\alpha} \quad (P=1,2,.., N; \,\,\, \alpha=1,2,..,N-1).
\eea 
%Under the gauge $U(1)$s, they transform as 
%\bea 
%\prod_\alpha U(1)_\alpha: \quad \phi_\alpha \to e^{ i \Lambda_\alpha} \phi_\alpha \quad (\alpha=1,\cdots,N-1),
%\quad \phi_N \to \phi_N.
%\eea 
Although it is not essential for our discussion, to make the model simpler, we assume the discrete symmetry $\mathbb{Z}_{q+1}$ under which
\bea 
\mathbb{Z}_{q+1}:\quad \theta^i\to \theta^i + (-1)^i \frac{2\pi}{q+1},\quad \phi_{P} \to e^{i (-1)^P(P+1) \frac{2\pi}{q+1}}\phi_{P}. 
\eea 
Then the scalar potentials of $\phi_P$ takes the form 
\bea \label{scalar_pot}
V(\theta^i, \phi_I)&=& \sum_P m_P^2 |\phi_P|^2 +  \sum_{PQ}  \lambda_{PQ}|\phi_P|^2 |\phi_Q|^2 
%(\textrm{$\theta$-independent operators with dim$>4$})
\nonumber\\
&& \hskip -2cm -\, 
 \epsilon_1 e^{i (q+1)\theta^1}\phi_1^{*q+1} - \epsilon_2 e^{i \theta^2} \phi_1^q \phi_2^*  -\cdots -\epsilon_N e^{i \theta^N} \phi_{N-1}^q \phi_N^* 
-\epsilon_{N+1}  \phi_N^{q+1} +  {\rm h.c.} + ... ,
\nonumber \eea
where the ellipsis denotes the higher-dimensional terms. Here we consider the case with $q=3$ or 2, but pretend that $q$
is a generic integer. 
% where $\epsilon_I$ ($I=1,2,.., N+1$) are assumed to be small enough   
%compared to the other model parameters.
%The first line of (\ref{scalar_pot}) 
%denotes $\theta$-independent combination of scalar fields.
%In the second line, the lowest order $\theta$-dependent operators are presented.
%One can find that the gauge invariance naturally 
%provides a clockworky mixing among $\theta^i$ and $\phi_I$. 
After integrating out $\zeta^\alpha$ eaten by the $U(1)_\alpha$ gauge fields, the second line of the potential becomes
\bea \label{pot_xi_scalar}
\hskip-0.2cm -\epsilon_1 e^{i b_1\xi/M_\xi} \phi_1^{* q+1} - 
\epsilon_2 e^{i b_2\xi/M_\xi} \phi_1^q \phi_2^* -\cdots - 
\epsilon_N e^{i b_N\xi/M_\xi} \phi_{N-1}^q \phi_N^* - \epsilon_{N+1} \phi_N^{q+1} 
+ {\rm h.c.},\quad 
\eea 
where 
\bea \label{coeff_ex1}
b_1 =\frac{(q+1)(q^2- 1)}{(q^{N} - q^{-N})q},\quad 
b_2= \frac{q^2-  1}{(q^N-q^{-N})q^2} ,\quad \cdots ,\quad  b_N =  \frac{q^2- 1}{(q^N-q^{-N})q^{N}}. 
\eea 
One can easily arrange the model to have non-zero vacuum values of $\phi_P$, and then $\phi_P$ can be decomposed as
\bea 
\phi_P(x) = \frac{1}{\sqrt{2}} \Big(v_P + h_P(x)\Big) e^{ i a_P(x)/v_P},
\eea 
where $v_P/\sqrt{2}=\langle \phi_P\rangle$ and $h_P$ denotes the radial fluctuation of $\phi_P$.
Here we are interested in the limit that $\epsilon_I$ are small enough, so that the phase fields $a_P$ can be regarded as light
pseudo-Goldstone bosons. Then the massive $h_P$ can be safely integrated out, while leaving the
following effective potential of $a_P$ and $\xi$:
\bea  \label{scalar_pot_N+1}
\hskip -0.7cm 
V_{\rm eff}(\xi, a_I) &=& -\Lambda_1^4 \cos \Big((q+1)\frac{a_1}{v_1} - 
b_1 \frac{\xi}{M_\xi} \Big)  - \Lambda_2^4\cos \Big(q\frac{a_1}{v_1} -\frac{a_2}{v_2} + 
 b_2 \frac{\xi}{M_\xi} \Big) +\cdots\nonumber\\
&& \hskip -0.5cm -\, \Lambda_N^4
\cos\Big(q\frac{a_{N-1}}{v_{N-1}} - \frac{a_N}{v_N} +  b_N\frac{\xi}{M_\xi}\Big) - \Lambda_{N+1}^4 \cos \Big( (q+1)\frac{a_N}{v_{N+1}}\Big), 
\eea  
where 
\bea 
\Lambda_1^4= \frac{\epsilon_1 v_1^{q+1}}{2^{(q-1)/2}},~ 
\Lambda_2^4 =\frac{\epsilon_2 v_1^q v_2}{2^{(q-1)/2}},~ \cdots,~ 
\Lambda_N^4 = \frac{\epsilon_N v_{N-1}^q v_N}{2^{(q-1)/2}},~ 
\Lambda_{N+1}^4 = \frac{\epsilon_{N+1} v_N^{q+1}}{2^{(q-1)/2}}.
\eea 
The above potential involves $N+1$ independent terms for the $N+1$ pseudo-Goldstone bosons involving $\xi$ and $a_P$, so can provide a non-trivial effective potential of $\xi$ which would be the lightest pseudo-Goldstone boson in the parameter limit
$v_P\ll M_\xi$. In fact, the above potential of $N+1$ pseudo-Goldstone bosons reveal the clockwork structure studied before \cite{Choi:2014rja,Choi:2015fiu,
Kaplan:2015fuy, Giudice:2016yja}. 
To proceed, let us first minimize the above potential {\it except} the first and last terms under the assumption that
all $\Lambda_I$ ($I=2,3,.., N$) are comparable to each other. This
 results in the following 
 $\xi$-dependent vacuum values of $a_I$ ($I=1,2,.., N-1$): 
\bea \label{eom_a_I}
\hskip -0.7cm
&&\frac{a_1}{v_1} =\frac{1}{q}\left(\frac{a_2}{v_2} - \frac{b_2\xi}{M_\xi}\right),~
\frac{a_2}{v_2} =\frac{1}{q}\left(\frac{a_3}{v_3} - \frac{b_3\xi}{M_\xi}\right),~ \cdots,\quad \frac{a_{N-1}}{v_{N-1}} =\frac{1}{q}\left(\frac{a_{N}}{v_{N}} -\frac{b_N\xi}{M_\xi}\right).\eea 
Inserting these to (\ref{scalar_pot_N+1}), we get the effective potential of $\xi$ and $a_N$, which is given by
\bea 
\label{two_axion}
V_{\rm eff}(\xi, a_N) =  -\Lambda_1^4\cos\left(\frac{q+1}{q^{N-1}}
\Big(\frac{a_N}{v_N} - \frac{\xi}{M_\xi}\Big)\right)-
\Lambda_{N+1}^4 \cos\Big((q+1)\frac{a_N}{v_{N}}\Big).
\eea 
If $\Lambda_1\sim \Lambda_{N+1}$, $a_N$ gets a mass dominantly from the second term with a vanishing vacuum value, which would result in
\bea
V_{\rm eff}(\xi)& \simeq& - \Lambda_1^4 \cos \Big(\frac{q+1}{q^{N-1}}\frac{\xi}{M_\xi}\Big).
\eea 
In this case the scalar potential (\ref{pot_xi_scalar}) not only provides a non-trivial effective potential of $\xi$,
but also enlarges the axion field range from $M_\xi$ to $q^{N-1}M_\xi$ through the clockwork mechanism \cite{Choi:2014rja,Choi:2015fiu,
Kaplan:2015fuy, Giudice:2016yja}.
Yet there exists a parameter limit where a non-trivial potential of $\xi$ is generated while keeping the axion field range as
$M_\xi$.  
%The reason that 
%we did not integrate out $a_N$ yet is that it can be  lighter than 
%other $a_I$ axions. 
If $\Lambda_{N+1}^2/\Lambda_1^2 \ll 1/q^{N-1}$, $a_N$ gets a mass dominantly from the first term of (\ref{two_axion}) with
the $\xi$-dependent vacuum value\bea
\frac{a_N}{v_N}=\frac{\xi}{M_\xi},\eea
yielding the effective potential
\bea 
V_{\rm eff}(\xi)
&\simeq & - \Lambda_{N+1}^4 \cos \Big((q+1)\frac{\xi}{M_\xi}\Big)  
\eea 
without changing the field range of $\xi$. At any rate,
our example shows again that $\xi$ is so well protected by $\prod_\alpha U(1)_\alpha$, so it requires a highly non-trivial engineering to generate  an effective potential of $\xi$ 
in the limit $N\gg 1$.

\section{Conclusion} \label{sec:con}

St\"uckelberg mixing between axions and $U(1)$ gauge bosons can result in  a variety of  interesting consequences in low energy axion physics. In this paper,
we studied those consequences 
%between axions and $U(1)$ gauge bosons
in a general framework which can be applied for many different situations.
More specifically we derived the field range of the gauge-invariant axion combination $\xi$ 
for generic form of axion kinetic metric  and $U(1)$ gauge charges, 
   %We note that in some cases the axion coupling to non-Abelian gauge fields,  $\xi G^{a\mu\nu}\tilde G^a_{\mu\nu}$, can have an %unambiguous connection  to the axion field range $\Delta\xi$ {\it only} when the coupling is shifted by an appropriate field %redefinition involving $\xi$-dependent $U(1)_A$ transformation of the gauge-charged fermion fields. 
 and examined the low energy axion couplings to matter and gauge fields in models with St\"uckelberg mixing, as well as some of their physical consequences.  
 St\"uckelberg mixing typically reduces the field range of  $\xi$  compared to the mass scales introduced in the UV theory.
In particular,
for the case of St\"uckelberg mixing between   $N$ axions and $(N-1)$ $U(1)$ gauge bosons in the limit $N\gg 1$,
the axion field range can be exponentially 
reduced relative to the mass scales in the UV theory.
As is well known, axion couplings in the effective lagrangian vary under the axion-dependent field redefinition of
matter fields, so depend on the choice of the matter field basis. It is noted that  
in some parameter limit of St\"uckelberg mixing the axion couplings to matter and gauge fields can have hierarchically different size in different field bases,  and then
one needs to consider the basis-independent (reparameterization-invariant) combination of couplings rather than focusing on a specific basis-dependent coupling
to see the physical consequence of the model.

\begin{acknowledgments}
	This work is supported by IBS under the project code, IBS-R018-D1.
\end{acknowledgments}


\begin{thebibliography}{99}

\bibitem{Kim:2008hd} 
  For reviews on axions, see for instance  J.~E.~Kim and G.~Carosi,
  %``Axions and the Strong CP Problem,''
  Rev.\ Mod.\ Phys.\  {\bf 82}, 557 (2010)
  [arXiv:0807.3125 [hep-ph]];
%\bibitem{Marsh:2015xka} 
M.~Kawasaki and K.~Nakayama,
  %``Axions: Theory and Cosmological Role,''
  Ann.\ Rev.\ Nucl.\ Part.\ Sci.\  {\bf 63}, 69 (2013)
  [arXiv:1301.1123 [hep-ph]];
  D.~J.~E.~Marsh,
  %``Axion Cosmology,''
  Phys.\ Rept.\  {\bf 643}, 1 (2016)
  [arXiv:1510.07633 [astro-ph.CO]].
 
\bibitem{Freese:1990rb} 
  K.~Freese, J.~A.~Frieman and A.~V.~Olinto,
  %``Natural inflation with pseudo - Nambu-Goldstone bosons,''
  Phys.\ Rev.\ Lett.\  {\bf 65}, 3233 (1990).
  
  \bibitem{Choi:1999xn} 
  K.~Choi,
  %``String or M theory axion as a quintessence,''
  Phys.\ Rev.\ D {\bf 62}, 043509 (2000)
  [hep-ph/9902292].
  
  \bibitem{Graham:2015cka} 
  P.~W.~Graham, D.~E.~Kaplan and S.~Rajendran,
  %``Cosmological Relaxation of the Electroweak Scale,''
  Phys.\ Rev.\ Lett.\  {\bf 115}, no. 22, 221801 (2015)
  [arXiv:1504.07551 [hep-ph]]. 
 
\bibitem{Witten:1984dg} 
  E.~Witten,
  %``Some Properties of O(32) Superstrings,''
  Phys.\ Lett.\  {\bf 149B}, 351 (1984). 
 
\bibitem{Choi:1985je} 
  K.~Choi and J.~E.~Kim,
  %``Harmful Axions in Superstring Models,''
  Phys.\ Lett.\  {\bf 154B}, 393 (1985)
  Erratum: [Phys.\ Lett.\  {\bf 156B}, 452 (1985)];
  K.~Choi and J.~E.~Kim,
  %``Compactification and Axions in E(8) x E(8)-prime Superstring Models,''
  Phys.\ Lett.\  {\bf 165B}, 71 (1985).
  
  \bibitem{Svrcek:2006yi} 
  P.~Svrcek and E.~Witten,
  %``Axions In String Theory,''
  JHEP {\bf 0606}, 051 (2006)
  [hep-th/0605206].
  
  \bibitem{Arvanitaki:2009fg} 
  A.~Arvanitaki, S.~Dimopoulos, S.~Dubovsky, N.~Kaloper and J.~March-Russell,
  %``String Axiverse,''
  Phys.\ Rev.\ D {\bf 81}, 123530 (2010)
  [arXiv:0905.4720 [hep-th]].
  
\bibitem{ArkaniHamed:2006dz} 
  N.~Arkani-Hamed, L.~Motl, A.~Nicolis and C.~Vafa,
  %``The String landscape, black holes and gravity as the weakest force,''
  JHEP {\bf 0706}, 060 (2007)
  [hep-th/0601001].  
 
\bibitem{ArkaniHamed:2003wu} 
  N.~Arkani-Hamed, H.~C.~Cheng, P.~Creminelli and L.~Randall,
  %``Extra natural inflation,''
  Phys.\ Rev.\ Lett.\  {\bf 90}, 221302 (2003)
  [hep-th/0301218]. 
 
 
 \bibitem{Kim:2004rp} 
  J.~E.~Kim, H.~P.~Nilles and M.~Peloso,
  %``Completing natural inflation,''
  JCAP {\bf 0501}, 005 (2005)
  [hep-ph/0409138].   
  

  \bibitem{Dimopoulos:2005ac} 
S.~Dimopoulos, S.~Kachru, J.~McGreevy and J.~G.~Wacker,
%``N-flation,''
JCAP {\bf 0808}, 003 (2008)
[hep-th/0507205].
  
 
  
  
  
  
 \bibitem{Silverstein:2008sg} 
  E.~Silverstein and A.~Westphal,
  %``Monodromy in the CMB: Gravity Waves and String Inflation,''
  Phys.\ Rev.\ D {\bf 78}, 106003 (2008)
  [arXiv:0803.3085 [hep-th]].
  
  \bibitem{Kaloper:2008fb} 
  N.~Kaloper and L.~Sorbo,
  %``A Natural Framework for Chaotic Inflation,''
  Phys.\ Rev.\ Lett.\  {\bf 102}, 121301 (2009)
  [arXiv:0811.1989 [hep-th]].
  
 \bibitem{Marchesano:2014mla} 
  F.~Marchesano, G.~Shiu and A.~M.~Uranga,
  %``F-term Axion Monodromy Inflation,''
  JHEP {\bf 1409}, 184 (2014)
  [arXiv:1404.3040 [hep-th]].
 
 %\cite{Choi:2014rja}
\bibitem{Choi:2014rja} 
  K.~Choi, H.~Kim and S.~Yun,
  %``Natural inflation with multiple sub-Planckian axions,''
  Phys.\ Rev.\ D {\bf 90}, 023545 (2014)
  [arXiv:1404.6209 [hep-th]].
  %%CITATION = doi:10.1103/PhysRevD.90.023545;%%
  %152 citations counted in INSPIRE as of 22 Sep 2019


\bibitem{Higaki:2014pja} 
  T.~Higaki and F.~Takahashi,
  %``Natural and Multi-Natural Inflation in Axion Landscape,''
  JHEP {\bf 1407}, 074 (2014)
  [arXiv:1404.6923 [hep-th]].

\bibitem{Bachlechner:2014hsa} 
  T.~C.~Bachlechner, M.~Dias, J.~Frazer and L.~McAllister,
  %``Chaotic inflation with kinetic alignment of axion fields,''
  Phys.\ Rev.\ D {\bf 91}, no. 2, 023520 (2015)
  [arXiv:1404.7496 [hep-th]].
  
  \bibitem{Ben-Dayan:2014zsa} 
  I.~Ben-Dayan, F.~G.~Pedro and A.~Westphal,
  %``Hierarchical Axion Inflation,''
  Phys.\ Rev.\ Lett.\  {\bf 113}, 261301 (2014)
  [arXiv:1404.7773 [hep-th]]. 
  
  
  %\cite{Rudelius:2014wla}
  \bibitem{Rudelius:2014wla} 
  T.~Rudelius,
  %``On the Possibility of Large Axion Moduli Spaces,''
  JCAP {\bf 1504}, 049 (2015)
  [arXiv:1409.5793 [hep-th]].
  %%CITATION = doi:10.1088/1475-7516/2015/04/049;%%
  %86 citations counted in INSPIRE as of 24 Sep 2019
  
    
  %\cite{Bachlechner:2014gfa}
  \bibitem{Bachlechner:2014gfa} 
  T.~C.~Bachlechner, C.~Long and L.~McAllister,
  %``Planckian Axions in String Theory,''
  JHEP {\bf 1512}, 042 (2015)
  [arXiv:1412.1093 [hep-th]].
  %%CITATION = doi:10.1007/JHEP12(2015)042;%%
  %59 citations counted in INSPIRE as of 23 Sep 2019
 
\bibitem{delaFuente:2014aca} 
  A.~de la Fuente, P.~Saraswat and R.~Sundrum,
  %``Natural Inflation and Quantum Gravity,''
  Phys.\ Rev.\ Lett.\  {\bf 114}, no. 15, 151303 (2015)
  [arXiv:1412.3457 [hep-th]].
  



  
  
 %\cite{Shiu:2015uva}
\bibitem{Shiu:2015uva} 
  G.~Shiu, W.~Staessens and F.~Ye,
  %``Widening the Axion Window via Kinetic and Stückelberg Mixings,''
  Phys.\ Rev.\ Lett.\  {\bf 115}, 181601 (2015)
  [arXiv:1503.01015 [hep-th]].
  %%CITATION = doi:10.1103/PhysRevLett.115.181601;%%
  %27 citations counted in INSPIRE as of 22 Sep 2019
 

 %\cite{Shiu:2015xda}
\bibitem{Shiu:2015xda} 
  G.~Shiu, W.~Staessens and F.~Ye,
  %``Large Field Inflation from Axion Mixing,''
  JHEP {\bf 1506}, 026 (2015)
  [arXiv:1503.02965 [hep-th]].
  %%CITATION = doi:10.1007/JHEP06(2015)026;%%
  %31 citations counted in INSPIRE as of 22 Sep 2019


%\cite{Montero:2015ofa}
\bibitem{Montero:2015ofa} 
M.~Montero, A.~M.~Uranga and I.~Valenzuela,
%``Transplanckian axions!?,''
JHEP {\bf 1508}, 032 (2015)
[arXiv:1503.03886 [hep-th]].
%%CITATION = doi:10.1007/JHEP08(2015)032;%%
%125 citations counted in INSPIRE as of 24 Sep 2019
 
  
  \bibitem{Brown:2015iha} 
  J.~Brown, W.~Cottrell, G.~Shiu and P.~Soler,
  %``Fencing in the Swampland: Quantum Gravity Constraints on Large Field Inflation,''
  JHEP {\bf 1510}, 023 (2015)
  [arXiv:1503.04783 [hep-th]]. 
  
  
  
   
  \bibitem{Hebecker:2015rya} 
  A.~Hebecker, P.~Mangat, F.~Rompineve and L.~T.~Witkowski,
  %``Winding out of the Swamp: Evading the Weak Gravity Conjecture with F-term Winding Inflation?,''
  Phys.\ Lett.\ B {\bf 748}, 455 (2015)
  [arXiv:1503.07912 [hep-th]].
 

\bibitem{Junghans:2015hba} 
D.~Junghans,
%``Large-Field Inflation with Multiple Axions and the Weak Gravity Conjecture,''
JHEP {\bf 1602}, 128 (2016)
[arXiv:1504.03566 [hep-th]]. 
  
  %\cite{Heidenreich:2015wga}
  \bibitem{Heidenreich:2015wga} 
  B.~Heidenreich, M.~Reece and T.~Rudelius,
  %``Weak Gravity Strongly Constrains Large-Field Axion Inflation,''
  JHEP {\bf 1512}, 108 (2015)
  [arXiv:1506.03447 [hep-th]].
  %%CITATION = doi:10.1007/JHEP12(2015)108;%%
  %121 citations counted in INSPIRE as of 24 Sep 2019
   
 
 
 \bibitem{Choi:2015fiu} 
  K.~Choi and S.~H.~Im,
  %``Realizing the relaxion from multiple axions and its UV completion with high scale supersymmetry,''
  JHEP {\bf 1601}, 149 (2016)
  [arXiv:1511.00132 [hep-ph]].
  
  \bibitem{Kaplan:2015fuy} 
  D.~E.~Kaplan and R.~Rattazzi,
  %``Large field excursions and approximate discrete symmetries from a clockwork axion,''
  Phys.\ Rev.\ D {\bf 93}, no. 8, 085007 (2016)
  [arXiv:1511.01827 [hep-ph]].
  
  \bibitem{Giudice:2016yja} 
  G.~F.~Giudice and M.~McCullough,
  %``A Clockwork Theory,''
  JHEP {\bf 1702}, 036 (2017)
  [arXiv:1610.07962 [hep-ph]].
 
 
 
 %\cite{Bachlechner:2017zpb}
 \bibitem{Bachlechner:2017zpb} 
 T.~C.~Bachlechner, K.~Eckerle, O.~Janssen and M.~Kleban,
 %``Multiple-axion framework,''
 Phys.\ Rev.\ D {\bf 98}, no. 6, 061301 (2018)
 [arXiv:1703.00453 [hep-th]].
 %%CITATION = doi:10.1103/PhysRevD.98.061301;%%
 %15 citations counted in INSPIRE as of 24 Sep 2019
 
 %\cite{Bachlechner:2017hsj}
 \bibitem{Bachlechner:2017hsj} 
 T.~C.~Bachlechner, K.~Eckerle, O.~Janssen and M.~Kleban,
 %``Systematics of Aligned Axions,''
 JHEP {\bf 1711}, 036 (2017)
 [arXiv:1709.01080 [hep-th]].
 %%CITATION = doi:10.1007/JHEP11(2017)036;%%
 %13 citations counted in INSPIRE as of 24 Sep 2019
 
 
 \bibitem{Shiu:2018unx} 
  G.~Shiu and W.~Staessens,
  %``Phases of Inflation,''
  JHEP {\bf 1810}, 085 (2018)
  [arXiv:1807.00888 [hep-th]].
 
%\bibitem{Agrawal:2017cmd} 
% P.~Agrawal, J.~Fan, M.~Reece and L.~T.~Wang,
 %``Experimental Targets for Photon Couplings of the QCD Axion,''
% JHEP {\bf 1802}, 006 (2018)
% [arXiv:1709.06085 [hep-ph]].
 
 %\cite{Fonseca:2019aux}
\bibitem{Fonseca:2019aux} 
  N.~Fonseca, B.~von Harling, L.~de Lima and C.~S.~Machado,
  %``The Super-Planckian Axion Strikes Back,''
  arXiv:1906.10193 [hep-ph].
  %%CITATION = ARXIV:1906.10193;%%
 
 
 
 
%\cite{Bonnefoy:2018ibr}
\bibitem{Bonnefoy:2018ibr} 
  Q.~Bonnefoy, E.~Dudas and S.~Pokorski,
  %``Axions in a highly protected gauge symmetry model,''
  Eur.\ Phys.\ J.\ C {\bf 79}, no. 1, 31 (2019)
  [arXiv:1804.01112 [hep-ph]].
  %%CITATION = doi:10.1140/epjc/s10052-018-6528-z;%%
  %6 citations counted in INSPIRE as of 22 Sep 2019 
 
 
 
 
 
%%%%%%%%%%%%%%%%%%%%%%%%Extras%%%%%%%%%%%%%%%%%%%%%%%%%%% 
 
 
 
 
%\bibitem{Saraswat:2016eaz} 
% P.~Saraswat,
% %{\it Weak gravity conjecture and effective field theory},
% Phys.\ Rev.\ D {\bf 95}, no. 2, 025013 (2017)
% %%doi:10.1103/PhysRevD.95.025013
% [arXiv:1608.06951 [hep-th]].
 
 %\cite{Agrawal:2017cmd}
 
 %%CITATION = doi:10.1007/JHEP02(2018)006;%%
 %28 citations counted in INSPIRE as of 23 Sep 2019
 

 
 
 
\end{thebibliography}
\end{document}